\documentclass[preprint,authoryear,3p,times,onecolumn]{elsarticle}

\usepackage{graphicx, color, amsthm, amssymb, caption}
\usepackage{txfonts, longtable, dcolumn, ctable, lscape, multirow, lineno, bbding, pifont}
\usepackage{natbib}
\bibpunct{(}{)}{;}{a}{}{,}
\journal{Icarus}

\usepackage{microtype}

\begin{document}

\begin{frontmatter}


\ead{hanus.home@gmail.com}
\cortext[cor1]{Corresponding author. Tel: +420 221912572. Fax: +420 221912577.}


\title{Sizes of main-belt asteroids by combining shape models and Keck adaptive aptics observations}

\author[1]{J.~Hanu\v s\corref{cor1}}
\author[2]{F.~Marchis}
\author[1]{J.~\v Durech}

\address[1]{Astronomical Institute, Faculty of Mathematics and Physics,
Charles University, V Hole\v sovi\v ck\' ach 2, 180 00 Prague, Czech Republic.}
\address[2]{SETI Institute, Carl Sagan Center, 189 Bernado Avenue, Mountain View CA 94043, USA}

\begin{abstract}
\textbf{We select 50 main-belt asteroids with a diameter between 20 and 400 km for which we have (i) shape models derived by the lightcurve inversion method (LI) and (ii) resolved observations of good quality collected with the Keck II adaptive optics (AO) system in the near-infrared. We derive the size of these asteroids by minimizing the difference between the contours from deconvolved AO images and the projected silhouettes calculated from the shape model at the time of the AO observations. We compute the volume-equivalent diameters for 48 of these asteroids. For 15 of them, we remove the ambiguity of the pole orientation typical for shape models derived by the LI. We have found that our equivalent diameters are smaller by 3\%, 7\%, and 2\% compared with the effective diameters derived from mid-IR photometric observations provided by IRAS, WISE and AKARI. For 40 asteroids with previously determined mass estimates, we compute their bulk densities and discuss the mass--density dependence with respect to taxonomic types.}
\end{abstract}


\end{frontmatter}

%\linenumbers

\section{Introduction}\label{introduction}

An important physical characteristic of an asteroid is its size. The measurement of an asteroid size in combination with its mass, allows us to directly compute the average density, and thus estimate its composition and the structure of its interior. The determination of the size of a small Solar System body is a difficult task to perform due to the small apparent size as observed from the Earth. For instance, at its best opposition, the angular diameter of the dwarf planet (1) Ceres is less than 0.7 arcsec, implying that high angular resolution instruments such as the Hubble Space Telescope or adaptive optics (AO) systems mounted on 8-10m class telescopes are necessary to directly image it.

Today, the mid-IR surveys of asteroids (IRAS, WISE, AKARI) are doubtlessly the most complete catalogs of size for small Solar System bodies. Based on the IRAS entire sky survey at wavelengths between 12 and 100 $\mu$m, \citet{Tedesco2002}'s catalog provides 2,228 observed asteroids, with diameters down to 7 km in the main belt. This represents just $\sim$0.5\% of the currently known $\sim$550,000 asteroids. The mid-IR survey on board the Japanese spacecraft AKARI gave an estimate of twice as many as the IRAS catalog (5,120 asteroids). A significant leap forward was made more recently with the NEOWISE catalog which provided from observations in 4 filter bands, the radiometric diameters of $\sim$100,000 main-belt asteroids \citep{Masiero2011}, so 1/5 of the cataloged ones. These measurements are however an estimate of the true size of the asteroids. They are based on assumption that the asteroid is spherical and thus the derived sizes are affected by a bias caused by the geometry of observations: the projection of the asteroid at the time of the thermal observation could be different by more than ten percent of the average projection (that corresponds to the spherical shape model). Additionally, the determination of the size is model-dependent and varies significantly with the thermal models which are being used. In \citet{Marchis2013a}, a comparison of the size estimate of (93) Minerva based on IRAS photometric measurements showed the model dependence on this analysis.

About three hundred convex shape models derived from the lightcurve inversion method \citep{Kaasalainen2001a,Kaasalainen2001b} are available in the Database of Asteroid Models from Inversion Techniques \citep[DAMIT,][\texttt{http://astro.troja.mff.cuni.cz/projects/asteroids3D}]{Durech2010}. This technique uses only disk-integrated photometry of asteroids to approximate three dimensional shape models built with convex polyhedrons. These shape models are not scaled in size because it is not possible to derive its size from the visible flux of the asteroid without having an accurate estimate of its albedo. Additionally, because of the symmetry of the lightcurve inversion method, two mirror solutions symmetrical in the ecliptic longitude of the pole direction by $\sim180^{\circ}$ are usually computed.

Size estimates of asteroids with an accuracy reaching $\sim$10\% can be determined by comparing the actual 2D projections of asteroid convex shape models with the stellar occultation measurements \citep{Timerson2009}. Using this approach, \citet{Durech2011} computed the sizes for 44 asteroids.

A more complex 3D shape-modeling technique called KOALA (Knitted Occultation, Adaptive optics, and Light\-curve Analysis) has been introduced recently by \citet{Carry2012}. This algorithm, based on multi-data set inversion \citep{Kaasalainen2011} and validated on asteroid (21) Lutetia, permits, in principle, a non-convex shape solution (e.g., if adaptive optics contours or stellar occultation measurements contain non-convex features). Similar analysis has been performed to derive the size and shape of (22) Kalliope \citep{Descamps2008}, (216) Kleopatra \citep{Descamps2011} and (93) Minerva \citep{Marchis2013a}.

By combining resolved direct images of asteroids collected with AO systems with their shape models derived by the lightcurve inversion method, we can infer the sizes of these asteroids \citep{Marchis2006,Carry2012}. By scaling the shape model to fit the estimated size of the resolved asteroid, we can derive a volume-equivalent diameter $D_{\mathrm{eq}}$, which is a diameter of a sphere of the same volume as the scaled convex model. Additionally, as shown in \citet{Marchis2006}, AO observations allow us to remove the uncertainty between two possible mirror solutions derived from lightcurve inversion method, and to also identify large surface non-convexities \citep[e.g., bilobated shape of (216) Kleopatra in][]{Descamps2011}.  

This work is based on the heritage of \citet{Marchis2006} and \citet{Carry2012} previous studies. We discuss in Section 2 a sample of AO observations and their corresponding convex shape models. We present in Section 3 the algorithm that we developed to derive the equivalent sizes of 48 asteroids and compare our results with thermal observations previously published. For five asteroids, we specifically developed new shape models. In Section 4 we estimate the bulk density of asteroids with a known estimated mass and conclude this work in Section 5 discussing the main outcomes of this study.

\section{Data}\label{sec:data}

\subsection{Adaptive Optics Observations}\label{sec:ao_data}

The W.M. Keck II telescope located atop Mauna Kea on the Big Island of Hawaii is equipped since 2000 with an AO system and the NIRC2 near-infrared camera. The AO system corrects in real-time Earth atmospheric turbulences providing an angular resolution close to the diffraction limit of the telescope at  $\sim$2.2 $\mu$m, so $\sim$45 mas (milliarcseconds) for bright targets (V$<$13.5) \citep{Wizinowich2000}. In 2007, the correction quality of the system was improved \citep{vanDam2004}, and the system is today capable of providing images with close to the diffraction limit of the telescope at shorter wavelength ($\sim$1.6 $\mu$m), hence with an angular resolution of 33 mas. Table~\ref{tab:ao_observations} lists all the resolved observations of asteroids collected in our observing programs using this instrument from 2005 to 2010. Before 2008, we recorded most of our data though a broad band Kp filter ($\lambda$ = 2.124 $\mu$m and $\Delta\lambda$ = 0.351 $\mu$m). After 2007, a narrow FeII band filter ($\lambda$ = 1.645 $\mu$m and $\Delta\lambda$ = 0.026 $\mu$m) were used to record most of the observations taking advantage of this improvement in image quality to resolve smaller asteroids and possibly detect closer and smaller companions (see for instance the detection of 3-km moons around (93) Minerva in \citet{Marchis2013a}.  

For the purpose of this work, we extracted a sample of asteroids from our set of 250 Keck AO observations of 164 asteroids based on a few criteria. We selected asteroids which have a convex shape model derived by the lightcurve inversion method and at least one disk-resolved image recorded at Keck telescope. The total number of asteroids in our sample is 50, distributed in 81 AO observations.

Each observation was performed and processed in a similar manner. The frames were recorded consecutively and co-added with a total exposure time of ~1 min per position using the narrowband camera with a pixel scale of 9.94 mas. A final image was obtained using our automatic pipeline while observing at the telescope by shift-adding 3--6 frames with an exposure time of 60 s (30 s x 2 co-adds). These frames were flat-field corrected, and we used a bad-pixel suppressing algorithm to improve the quality before shift-adding them. After applying this basic data processing, the final images reveal the resolved shape of the asteroid which varies in angular size from 100 to 300 mas in our sample. If  something suspicious was detected during the observing nights (presence of a possible companion, elongated and bilobated shape), additional observations were taken in Kp filter and processed in a similar way. Several times per night, we also observe an unresolved bright star to estimate the point spread function (PSF) of the AO system. This additional set of observations is useful to estimate the quality of the data, check for possible artefacts in the PSF of the instrument and deconvolve the data a posteriori. 

\begin{figure}
	\begin{center}
	 \resizebox{\hsize}{!}{\includegraphics{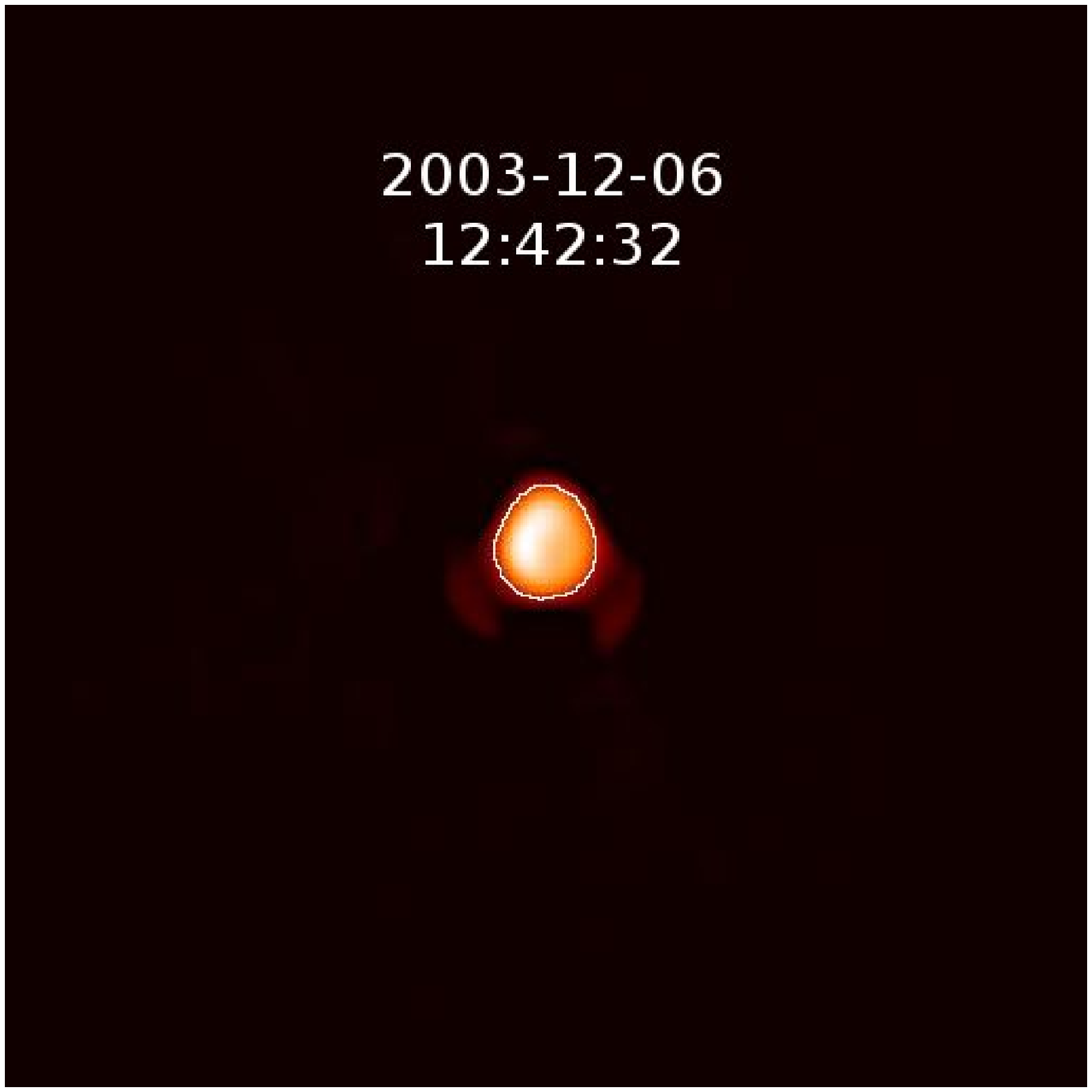}\,\,\,\includegraphics{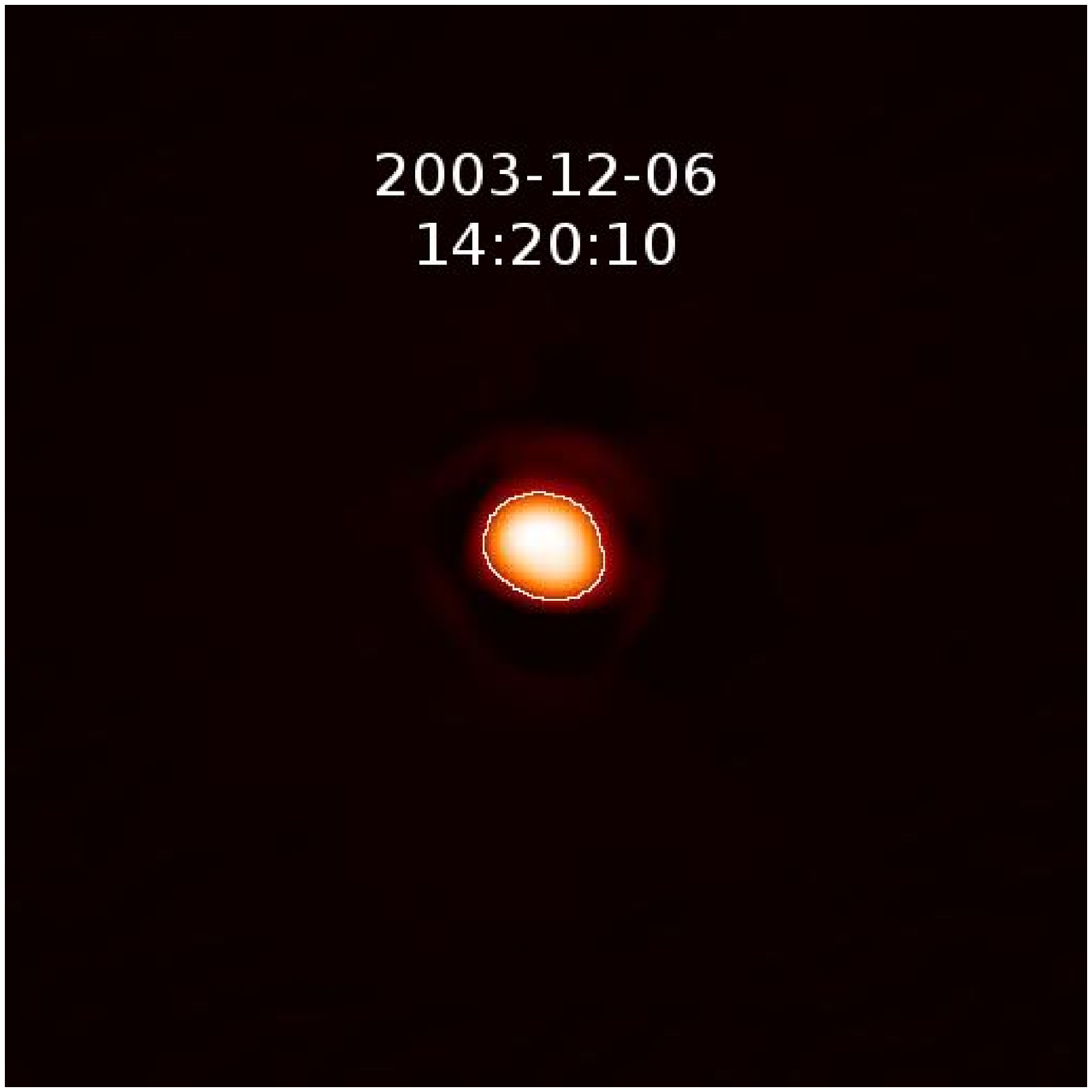}\,\,\,\includegraphics{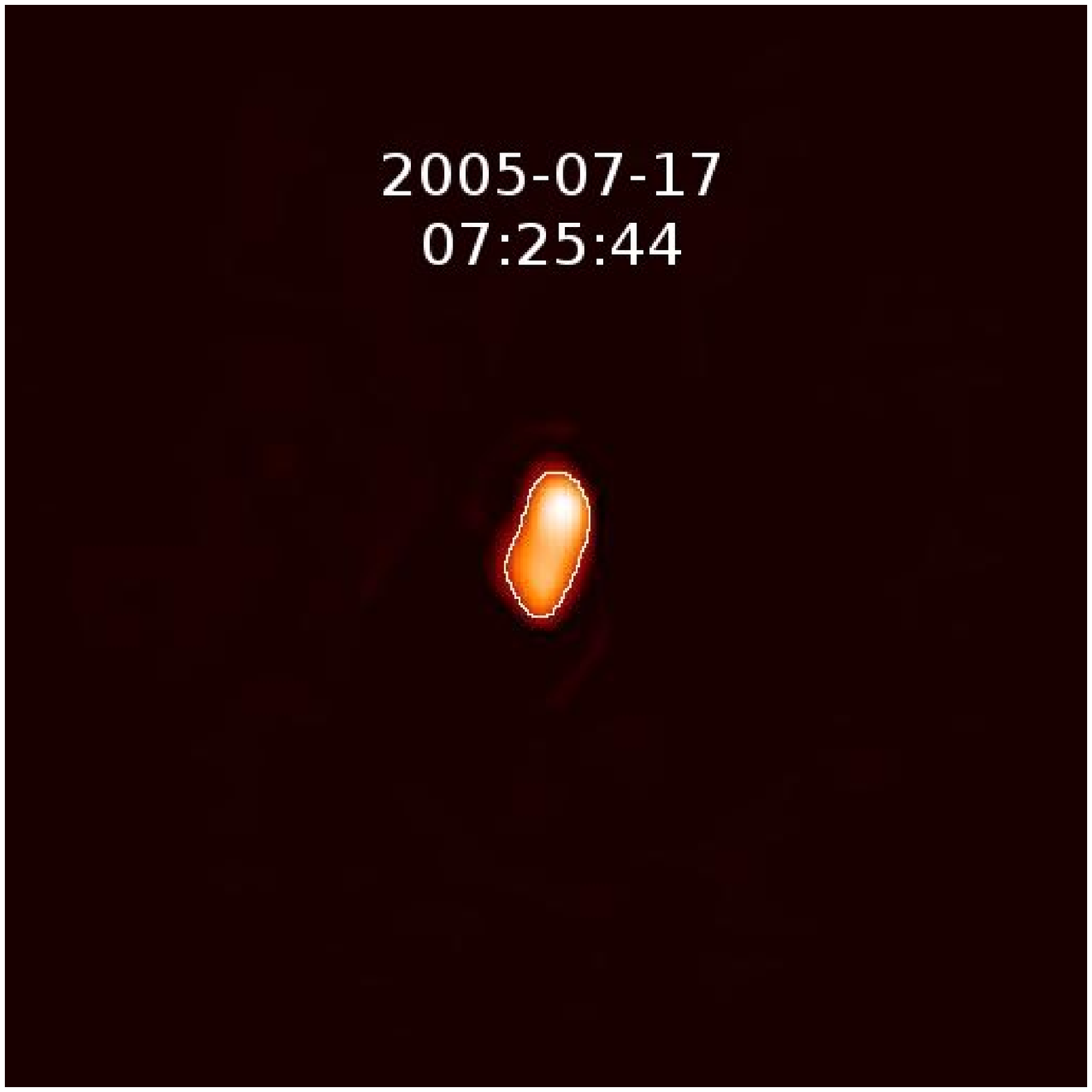}\,\,\,\includegraphics{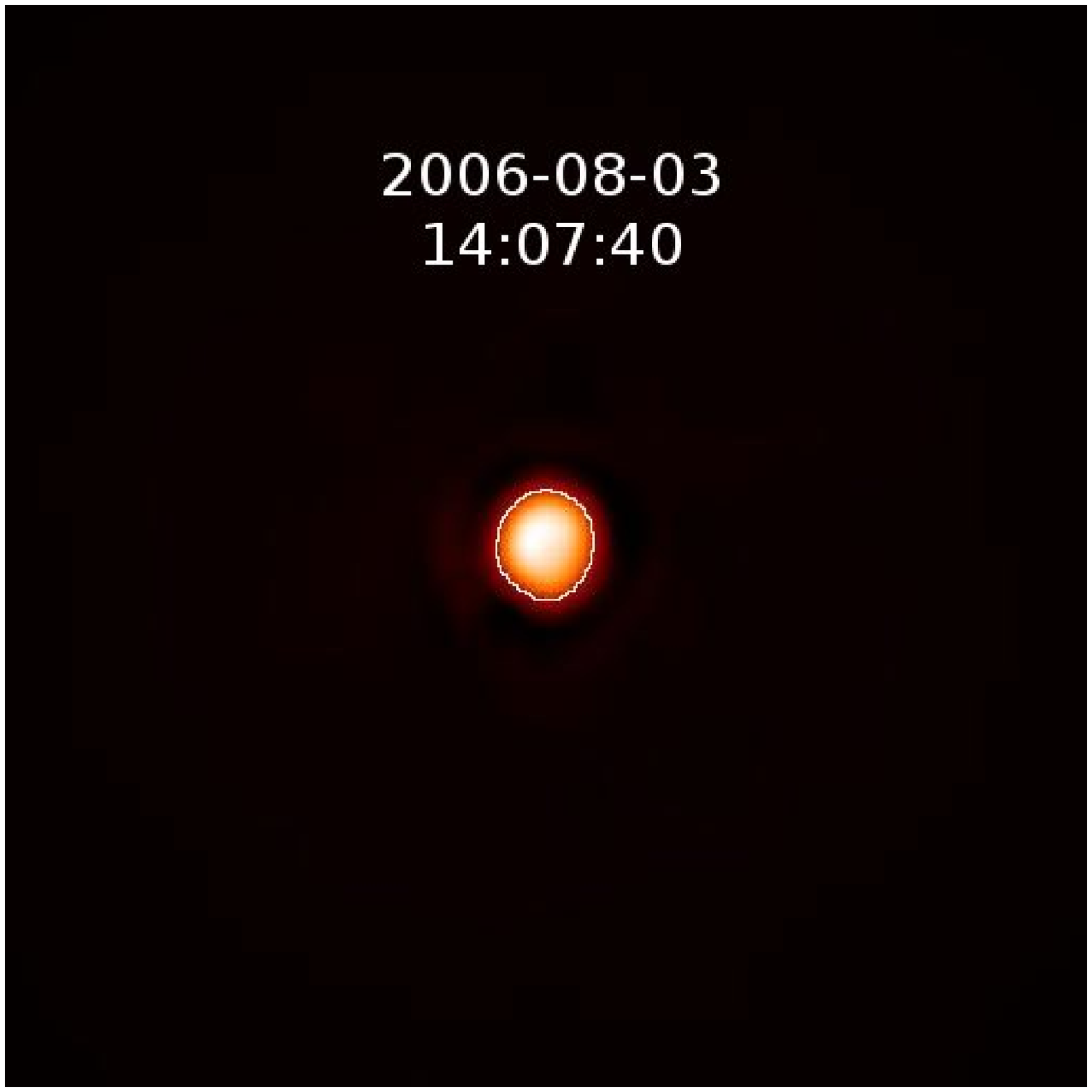}}
	\end{center}
	\caption{\label{img:ao_eugenia}Four adaptive optics images of asteroid (45)~Eugenia, contours were computed by the AIDA deconvolution algorithm \citep{Marchis2006, Hom2007}.}
\end{figure}

Since the final images have a high signal-to-noise ratio above 1000, and the Keck AO correction is relatively stable for bright V$\sim$11--13 targets, so we could apply the AIDA myopic deconvolution algorithm \citep{Hom2007} to improve the sharpness of the images, hence the estimation of the size and shape. We use as an initial guess for the PSF, a set of PSF frames collected during the night of observations through the same filter. From simulations of asteroid observations, \citet{Marchis2006} and \citet{Marchis2012a} showed that the typical error on the major-axis estimate of an asteroid resolved with $\sim$3 elements of resolution (SNR $\sim$2000, FWHM (PSF) $\sim$40 mas) is 3\%, which corresponds to 4 mas. Without deconvolution, the error is typically 7--10\% varying with the SNR and the quality of the PSF. The a posteriori deconvolution process improves the image quality and also the accuracy in determining the profile of the asteroid, hence its size. In Figure~\ref{img:ao_eugenia}, we show an example of four adaptive optics images of asteroid (45)~Eugenia. The silhouettes were computed with the AIDA deconvolution algorithm.

\onecolumn

\scriptsize{
\begin{longtable}{r@{\,\,\,}l ccccc}
\caption{\label{tab:ao_observations}List of resolved asteroid observations and their observational circumstances collected at the Keck II telescope and its adaptive optics from 2005 to 2010 selected for our study.}\\
\hline 
\multicolumn{2}{c} {Asteroid} & \multicolumn{1}{c} {Date} & \multicolumn{1}{c} {Time} & \multicolumn{1}{c} {Exposure} & \multicolumn{1}{c} {Filter} & \multicolumn{1}{c} {Airmass}\\
\multicolumn{2}{l} { } & [UT] & [UT] & [s] &  & \\
\hline\hline

\endfirsthead
\caption{continued.}\\

\hline
\multicolumn{2}{c} {Asteroid} & \multicolumn{1}{c} {Date} & \multicolumn{1}{c} {Time} & \multicolumn{1}{c} {Exposure} & \multicolumn{1}{c} {Filter} & \multicolumn{1}{c} {Airmass}\\
\multicolumn{2}{l} { } & [UT] & [UT] & [s] &  & \\
\hline\hline
\endhead
\hline
\endfoot

5 & Astraea & 17-VII-2005 & 11:36:36 & 163 & Kp & 1.216 \\
6 & Hebe & 28-VI-2010 & 13:07:33 & 180 & FeII & 1.446 \\
6 & Hebe & 29-XI-2010 & 07:10:05 & 180 & FeII & 1.333 \\
7 & Iris & 17-VII-2005 & 07:56:20 & 144 & Kp & 1.373 \\
7 & Iris & 16-VIII-2009 & 07:51:47 & 180 & FeII & 1.316 \\
7 & Iris & 16-VIII-2009 & 08:17:09 & 90 & FeII & 1.364 \\
8 & Flora & 28-VI-2010 & 12:24:54 & 180 & FeII & 1.647 \\
8 & Flora & 30-XI-2010 & 05:23:06 & 168 & FeII & 1.170 \\
9 & Metis & 25-X-2004 & 06:03:04 & 180 & Kp & 1.286 \\
9 & Metis & 25-X-2004 & 08:01:29 & 90 & Kp & 1.209 \\
10 & Hygiea & 19-IX-2008 & 13:49:10 & 180 & FeII & 1.104 \\
14 & Irene & 17-VII-2005 & 07:11:32 & 180 & Kp & 1.287 \\
14 & Irene & 17-VII-2005 & 07:17:50 & 49 & Kp & 1.303 \\
16 & Psyche & 16-VIII-2009 & 08:55:10 & 180 & FeII & 1.264 \\
19 & Fortuna & 16-VIII-2009 & 14:06:37 & 180 & FeII & 1.317 \\
22 & Kalliope & 12-XII-2006 & 13:41:59 & 180 & Kp & 1.285 \\
23 & Thalia & 16-VIII-2009 & 12:10:24 & 180 & FeII & 1.237 \\
23 & Thalia & 16-VIII-2009 & 12:17:14 & 180 & Kcont & 1.231 \\
28 & Bellona & 3-IV-2007 & 13:33:37 & 90 & Kp & 1.253 \\
29 & Amphitrite & 28-VI-2010 & 10:50:53 & 180 & FeII & 1.626 \\
30 & Urania & 28-VI-2010 & 08:07:20 & 180 & FeII & 1.398 \\
34 & Circe & 28-VI-2010 & 09:39:02 & 180 & FeII & 1.441 \\
37 & Fides & 16-VIII-2009 & 07:07:21 & 180 & FeII & 1.526 \\
39 & Laetitia & 17-VII-2005 & 11:04:02 & 90 & Kp & 1.146 \\
39 & Laetitia & 29-XI-2010 & 05:56:25 & 144 & FeII & 1.180 \\
40 & Harmonia & 2-VIII-2007 & 10:10:26 & 180 & BrG & 1.354 \\
40 & Harmonia & 11-XI-2011 & 12:01:40 & 180 & FeII & 1.120 \\
41 & Daphne & 30-XI-2010 & 09:17:42 & 180 & FeII & 1.067 \\
42 & Isis & 17-VII-2005 & 10:55:04 & 96 & Kp & 1.545 \\
45 & Eugenia & 6-XII-2003 & 12:42:32 & 45 & Kp & 1.067 \\
45 & Eugenia & 6-XII-2003 & 14:20:10 & 180 & Kp & 1.009 \\
45 & Eugenia & 17-VII-2005 & 07:25:44 & 180 & Kp & 1.227 \\
45 & Eugenia & 3-VIII-2006 & 14:07:40 & 180 & Kp & 1.098 \\
45 & Eugenia & 9-IX-2007 & 14:31:52 & 720 & Kp & 1.171 \\
45 & Eugenia & 9-IX-2007 & 15:20:50 & 180 & Kp & 1.048 \\
45 & Eugenia & 19-X-2007 & 12:05:30 & 900 & Ks & 1.206 \\
45 & Eugenia & 19-X-2007 & 12:54:38 & 360 & H & 1.067 \\
45 & Eugenia & 19-X-2007 & 13:29:59 & 180 & Ks & 1.020 \\
45 & Eugenia & 13-XII-2007 & 07:59:08 & 540 & Kp & 1.194 \\
45 & Eugenia & 28-VI-2010 & 09:52:28 & 60 & FeII & 1.560 \\
45 & Eugenia & 28-VI-2010 & 11:40:03 & 60 & FeII & 1.220 \\
45 & Eugenia & 29-XI-2010 & 05:10:14 & 60 & FeII & 1.617 \\
52 & Europa & 7-XII-2003 & 07:49:30 & 180 & Kp & 1.060 \\
54 & Alexandra & 28-VI-2010 & 13:20:31 & 180 & FeII & 1.336 \\
54 & Alexandra & 29-XI-2010 & 06:05:09 & 180 & FeII & 1.027 \\
68 & Leto & 3-VIII-2006 & 14:21:05 & 180 & Kp & 1.172 \\
69 & Hesperia & 2-VIII-2007 & 09:49:39 & 180 & Br$_\gamma$ & 1.141 \\
72 & Feronia & 17-VII-2005 & 11:12:21 & 149 & Kp & 1.124 \\
%79 & Eurynome & 19-IX-2008 & 05:52:46 & 180 & H & 1.412 \\
80 & Sappho & 2-VIII-2007 & 09:00:46 & 180 & Br$_\gamma$ & 1.098 \\
80 & Sappho & 28-VI-2010 & 06:56:39 & 180 & FeII & 1.163 \\
85 & Io & 2-VIII-2007 & 07:12:49 & 218 & Kp & 1.130 \\
87 & Sylvia & 25-X-2004 & 06:28:31 & 180 & Kp & 1.360 \\
87 & Sylvia & 12-XII-2006 & 16:07:34 & 360 & Kp & 1.250 \\
88 & Thisbe & 16-VIII-2009 & 09:21:04 & 180 & FeII & 1.188 \\
89 & Julia & 16-VIII-2009 & 12:53:03 & 180 & FeII & 1.025 \\
97 & Klotho & 17-VII-2005 & 10:33:59 & 180 & Kp & 1.136 \\
107 & Camilla & 25-X-2004 & 06:52:46 & 180 & Kp & 1.121 \\
107 & Camilla & 16-VIII-2009 & 06:46:42 & 180 & FeII & 1.187 \\
107 & Camilla & 28-VI-2010 & 10:19:15 & 180 & FeII & 1.937 \\
129 & Antigone & 28-VI-2010 & 07:18:33 & 180 & FeII & 1.160 \\
130 & Elektra & 7-XII-2003 & 07:16:10 & 180 & Kp & 1.428 \\
130 & Elektra & 15-I-2005 & 12:25:31 & 180 & Kp & 1.027 \\
130 & Elektra & 15-I-2005 & 14:14:01 & 180 & Kp & 1.081 \\
135 & Hertha & 19-IX-2008 & 13:27:45 & 180 & FeII & 1.020 \\
146 & Lucina & 17-VII-2005 & 08:12:50 & 180 & Kp & 1.407 \\
146 & Lucina & 28-VI-2010 & 10:31:12 & 180 & FeII & 2.619 \\
152 & Atala & 19-IX-2008 & 08:22:51 & 180 & FeII & 1.449 \\
165 & Loreley & 25-X-2004 & 09:03:31 & 180 & Kp & 1.004 \\
165 & Loreley & 29-XI-2010 & 10:01:51 & 180 & FeII & 1.073 \\
184 & Dejopeja & 28-VI-2010 & 08:39:41 & 180 & FeII & 1.870 \\
201 & Penelope & 28-VI-2010 & 06:44:00 & 180 & FeII & 1.302 \\
230 & Athamantis & 28-VI-2010 & 06:27:31 & 180 & FeII & 1.334 \\
250 & Bettina & 19-IX-2008 & 08:05:56 & 180 & FeII & 1.454 \\
276 & Adelheid & 30-VI-2005 & 14:32:05 & 90 & Kp & 1.053 \\
349 & Dembowska & 28-VI-2010 & 07:04:39 & 180 & FeII & 1.329 \\
354 & Eleonora & 2-VIII-2007 & 09:11:34 & 180 & Br$_\gamma$ & 1,197 \\
409 & Aspasia & 17-VII-2005 & 07:43:25 & 180 & Kp & 1.300 \\
409 & Aspasia & 17-VII-2005 & 07:47:47 & 90 & Kp & 1.312 \\
409 & Aspasia & 28-VI-2010 & 12:54:51 & 180 & FeII & 1.271 \\
423 & Diotima & 6-XII-2003 & 06:01:37 & 180 & Kp & 1.112 \\
471 & Papagena & 16-VIII-2009 & 05:54:14 & 180 & FeII & 1.402 \\
\hline
\end{longtable}
%\tablefoot{.}
}
\normalsize
%\twocolumn
%\end{landscape}
%\twocolumn

\subsection{Convex shape models of asteroids}\label{sec:models}

All 50 convex shape models used in this study were derived by the lightcurve inversion method and were based either on dense photometric data (26 of them) or dense data combined with sparse-in-time photometry from astrometric surveys (24 of them), and are available in the DAMIT database. 

In Table~\ref{tab:models}, we list for all these asteroid models including their parameters of rotational state, the number of used dense lightcurves observed during $N_{\mathrm{app}}$ apparitions, and sparse data points together with the reference. The uncertainty of the rotational period determination which is listed in Table~\ref{tab:models} is of the order of the last decimal place of period value $P$. The typical error for the orientation of the pole is (5--10$^{\circ}$)/$\cos \beta$ in longitude $\lambda$ and 5--20$^{\circ}$ in latitude $\beta$ (both uncertainties depend on the amount, timespan, variety and quality of used photometry). Most models which are based purely on dense photometry were published in \citet{Kaasalainen2002b} and \citet{Torppa2003}. These models are typically derived from a large number ($\sim$30--50) of individual dense lightcurves observed during $\sim$5--10 apparitions, and thus the uncertainties of parameters of the rotational state correspond to lower values of the aforementioned range. Models based on combined dense and sparse data were published in \citet{Hanus2011, Hanus2013a} and \citet{Durech2011}. Thanks to the sparse-in-time data that typically cover the time period of $\sim$15 years and $\sim$10 apparitions, $\sim$5--20 individual dense lightcurves from $\sim$3--7 apparitions were usually sufficient for a successful model determinations. The lower number of dense lightcurves is compensated by the sparse-in-time photometric data that cover various observational geometries. Due to the poor photometric quality of the sparse data, the uncertainties of parameters of the rotational state are higher than those of models based only on dense photometry.

\onecolumn

\scriptsize{
\begin{longtable}{r@{\,\,\,}l rrrr D{.}{.}{6} rrrc}
\caption{\label{tab:models}Rotational states and lightcurve information for 50 asteroid shape models derived from disk-integrated photometry by the lightcurve inversion method, for which we have quality AO images. For each asteroid, the table gives the ecliptic coordinates $\lambda_1$ and $\beta_1$ of the pole solution, the corresponding mirror solution $\lambda_2$ and $\beta_2$ (if any), the sidereal rotational period $P$, the number of dense lightcurves $N_{\mathrm{rel}}$ observed during $N_{\mathrm{app}}$ apparitions, the number of sparse data points $N_{\mathrm{sp}}$, and the references to the convex models.}\\
\hline 
\multicolumn{2}{c} {Asteroid} & \multicolumn{1}{c} {$\lambda_1$} & \multicolumn{1}{c} {$\beta_1$} & \multicolumn{1}{c} {$\lambda_2$} & \multicolumn{1}{c} {$\beta_2$} & \multicolumn{1}{c} {$P$} & $N_{\mathrm{rel}}$ & $N_{\mathrm{app}}$ & $N_{\mathrm{sp}}$ & \multicolumn{1}{c} {Reference}\\
\multicolumn{2}{l} { } & [deg] & [deg] & [deg] & [deg] & \multicolumn{1}{c} {[hours]} &  &  &  & \\
\hline\hline

\endfirsthead
\caption{continued.}\\

\hline
\multicolumn{2}{c} {Asteroid} & \multicolumn{1}{c} {$\lambda_1$} & \multicolumn{1}{c} {$\beta_1$} & \multicolumn{1}{c} {$\lambda_2$} & \multicolumn{1}{c} {$\beta_2$} & \multicolumn{1}{c} {$P$} & $N_{\mathrm{rel}}$ & $N_{\mathrm{app}}$  & $N_{\mathrm{sp}}$ & \multicolumn{1}{c} {Reference}\\
\multicolumn{2}{l} { } & [deg] & [deg] & [deg] & [deg] & \multicolumn{1}{c} {[hours]} &  &  &  & \\
\hline\hline
\endhead
\hline
\endfoot

5 & Astraea & 126 & 40 &  &  & 16.80061 & 24 & 7 & 153 & \citet{Durech2009} \\
6 & Hebe & 340 & 42 &  &  & 7.274471 & 39 & 14 &   & \citet{Torppa2003} \\
7 & Iris & 16 & 15 & 196 & 2 & 7.138843 & 31 & 11 & 629 & \citet{Durech2011} \\
8 & Flora & 155 & 6 & 335 & $-$5 & 12.86667 & 47 & 12 &  & \citet{Torppa2003} \\
9 & Metis & 180 & 22 &  &  & 12.86667 & 34 & 13 &  & \citet{Torppa2003} \\
10 & Hygiea & 312 & $-$42 & 122 & $-$44 & 27.6591 & 23 & 9 & 718 & \citet{Hanus2011} \\
14 & Irene & 95 & $-$11 & 271 & $-$12 & 15.02986 & 29 & 9 & 501 & \citet{Hanus2011} \\
15 & Eunomia & 3 & $-$67 &  &  & 6.082753 & 48 & 14 &  & \citet{Kaasalainen2002b}\\
16 & Psyche & 32 & $-$7 & 213 & 0 & 4.195948 & 114 & 18 &  & \citet{Kaasalainen2002b} \\
%17 & Thetis & 236 & 19 &  &  & 12.26603 & 52 & 8 & 221 & \citet{Durech2009} \\
19 & Fortuna & 98 & 57 &  &  & 7.44322 & 38 & 8 &  & \citet{Torppa2003} \\
22 & Kalliope & 196 & 3 &  &  & 4.148200 & 38 & 13 &  & \citet{Kaasalainen2002b} \\
23 & Thalia & 159 & $-$45 & 343 & $-$69 & 12.31241 & 45 & 11 &  & \citet{Torppa2003} \\
28 & Bellona & 282 & 6 & 102 & $-$8 & 15.70785 & 23 & 7 & 130 & \citet{Durech2011} \\
29 & Amphitrite & 138 & $-$21 &  &  & 5.390119 & 28 & 10 &  & \citet{Kaasalainen2002b} \\
30 & Urania & 107 & 23 & 284 & 20 & 13.68717 & 11 & 3 & 106 & \citet{Durech2009} \\
34 & Circe & 94 & 35 & 275 & 51 & 12.17458 & 16 & 5 & 114 & \citet{Durech2009} \\
37 & Fides & 270 & 19 & 89 & 27 & 7.332527 & 23 & 5 & 497 & \citet{Hanus2011} \\
39 & Laetitia & 323 & 32 &  &  & 5.138238 & 56 & 20 &  & \citet{Kaasalainen2002b} \\
40 & Harmonia & 22 & 31 & 206 & 39 & 8.908483 & 19 & 6 & 654 & \citet{Hanus2011} \\
41 & Daphne & 198 & $-$32 &  &  & 5.98798 & 23 & 7 &  & \citet{Kaasalainen2002b} \\
42 & Isis & 106 & 40 & 302 & 28 & 13.58364 & 28 & 7 & 511 & \citet{Hanus2011} \\
45 & Eugenia & 123 & $-$33 &  &  & 5.69914 & 44 & 8 &  & \citet{Kaasalainen2002b}\\
52 & Europa & 251 & 35 &  &  & 5.62996 & 49 & 11 &  & \citet{Kaasalainen2002b} \\
54 & Alexandra & 156 & 13 & 318 & 23 & 7.02264 & 25 & 6 & 144 & \citet{Warner2008a} \\
68 & Leto & 103 & 43 & 290 & 23 & 14.84547 & 12 & 2 & 441 & \citet{Hanus2011} \\
69 & Hesperia & 250 & 17 & 71 & $-$2 & 5.655340 & 35 & 7 & 397 & \citet{Hanus2011} \\
72 & Feronia & 287 & $-$39 & 102 & $-$55 & 8.09068 & 20 & 5 & 447 & \citet{Hanus2013a} \\
%79 & Eurynome & 228 & 30 & 54 & 24 & 5.97772 & 36 & 4 & 428 & \citet{Hanus2013a} \\
80 & Sappho & 194 & $-$26 &  &  & 14.03087 & 12 & 4 & 125 & \citet{Durech2009} \\
85 & Io & 95 & $-$65 &  &  & 6.874783 & 29 & 5 &  & \citet{Durech2011} \\
87 & Sylvia & 71 & 66 &  &  & 5.18364 & 32 & 7 &  & \citet{Kaasalainen2002b} \\
88 & Thisbe & 72 & 60 & 247 & 50 & 6.04131 & 20 & 6 &  & \citet{Torppa2003} \\
89 & Julia & 8 & $-$13 &  &  & 11.38834 & 13 & 2 & 339 & \citet{Durech2011} \\
97 & Klotho & 359 & 30 & 161 & 40 & 35.2510 & 25 & 6 & 542 & \citet{Hanus2011} \\
107 & Camilla & 73 & 54 &  &  & 4.843928 & 29 & 9 &  & \citet{Torppa2003} \\
129 & Antigone & 207 & 58 &  &  & 4.95715 & 34 & 10 &  & \citet{Torppa2003} \\
130 & Elektra & 64 & $-$88 &  &  & 5.22466 & 49 & 11 &  & \citet{Marchis2006} \\
135 & Hertha & 272 & 52 &  &  & 8.40060 & 42 & 8 &  & \citet{Torppa2003} \\
146 & Lucina & 139 & $-$14 & 305 & $-$41 & 18.5540 & 22 & 4 & 125 & \citet{Durech2009} \\
152 & Atala & 347 & 47 &  &  & 6.24472 & 2 & 1 & 101 & \citet{Durech2009} \\
165 & Loreley & 174 & 29 &  &  & 7.22439 & 29 & 6 & 201 & \citet{Durech2011} \\
184 & Dejopeja & 200 & 52 & 18 & 54 & 6.44111 & 17 & 6 &  & \citet{Marciniak2007} \\
201 & Penelope & 84 & $-$15 & 262 & $-$1 & 3.74745 & 32 & 7 &  & \citet{Torppa2003} \\
230 & Athamantis & 74 & 27 & 237 & 29 & 23.9845 & 36 & 7 &  & \citet{Torppa2003} \\
%233 & Asterope & 290 & 60 & 119 & 40 & 19.69802 & 13 & 3 & 445 & New model \\
250 & Bettina & 100 & 17 & 282 & $-$12 & 5.05442 & 23 & 6 &  & \citet{Torppa2003} \\
276 & Adelheid & 9 & $-$4 & 199 & $-$20 & 6.31920 & 31 & 7 &  & \citet{Marciniak2007} \\
349 & Dembowska & 149 & 41 & 322 & 18 & 4.70120 & 40 & 9 &  & \citet{Torppa2003} \\
354 & Eleonora & 144 & 54 &  &  & 4.277186 & 37 & 9 & 533 & \citet{Hanus2011} \\
409 & Aspasia & 3 & 30 &  &  & 9.02144 & 16 & 7 & 123 & \citet{Warner2008a} \\
423 & Diotima & 351 & 4 &  &  & 4.77538 & 50 & 11 &  & \citet{Marchis2006} \\
471 & Papagena & 223 & 67 &  &  & 7.11539 & 13 & 2 & 680 & \citet{Hanus2011} \\
%512 & Taurinensis & 324 & 45 &  &  & 5.58203 & 11 & 2 & 235 & \citet{Hanus2013a} \\
%675 & Ludmilla & 49 & 74 & 196 & 49 & 7.71549 & 38 & 5 &  & \citet{Torppa2003} \\

\hline
\end{longtable}
%\tablefoot{.}
}
\normalsize
%\twocolumn
%\end{landscape}
%\twocolumn

\section{Analysis of the data}\label{sec:analysis}

\subsection{Sizes from combining shape models and Keck adaptive optics images}\label{sec:ao_sizes}

We define the fundamental plane that passes through the center of the asteroid and is perpendicular to the line connecting the observer and the center of the asteroid, and also the coordinate system ($\xi$,~$\eta$) on the fundamental plane in the same way as \citet{Durech2011}. The coordinate system is centered at the center of the AO image. We extract the asteroid silhouette defined by pixel coordinates ($\xi_j$,~$\eta_j$)$_{\mathrm{AO}}$ directly from the AO observation by the AIDA deconvolution algorithm (see Figure~\ref{img:ao_eugenia}). While the distance $L$ of the asteroid from the Earth at the time of its observation with the AO system and the pixel scale of the image are known, we express the contour coordinates directly in kilometers. The distance $\gamma$ in mas on the AO image corresponds to the distance $x$ in kilometers by a relation:

\begin{equation}\label{eq:mas2km}
 x\mathrm{[km]} = \frac{\pi L\mathrm{[km]}}{180\cdot10^3\cdot3600} \gamma\mathrm{[mas]}.
\end{equation}

\begin{figure}
	\begin{center}
	 \resizebox{0.75\hsize}{!}{\includegraphics{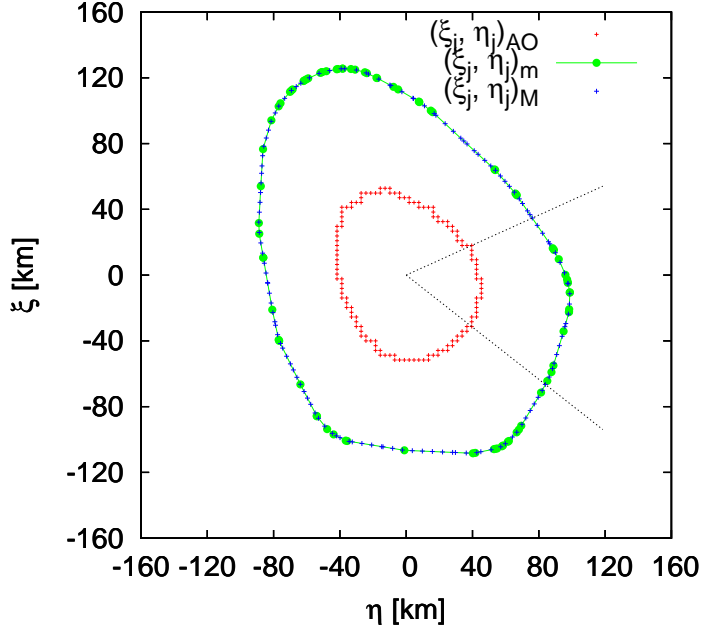}}\\
	 \end{center}
	 \caption{\label{img:intersections}Scaled contour ($\xi_j$,~$\eta_j$)$_{\mathrm{AO}}$ of the disk-resolved image  of asteroid (201)~Penelope (defined by pixels, red crosses), unscaled contour ($\xi_j$,~$\eta_j$)$_{\mathrm{m}}$ of the corresponding convex shape model computed for the time of the AO observation (green full circles connected with lines) and intersections ($\xi_j$,~$\eta_j$)$_{\mathrm{M}}$ (blue crosses) between the model contour and the lines which go from the center of the AO contour through the points ($\xi_j$,~$\eta_j$)$_{\mathrm{AO}}$ (two such lines are displayed by thin dotted lines). East points to the left and north up.}
\end{figure}

The convex hull of projected vertices (only both illuminated by the Sun and visible from the Earth) of the convex polyhedron onto the fundamental plane represents a silhouette of the model given by the coordinates ($\xi_j$,~$\eta_j$)$_{\mathrm{m}}$ (this is a valid approach for phase angles lower than 90$^{\circ}$, our typical observations are for phase angles lower than 30$^{\circ}$). The model contour is measured from the convex shape model orientation at the time of the AO observation, however, we correctly account for the light-time effect. To determine the true size of the model, we minimize the difference between these two silhouettes. To successfully compare both contours, we find the points on the model contour that correspond to the points defining the AO contour. These points lie on the intersections between the model contour and the lines which go from the center of the AO contour (i.e., center of the coordinate system) through the points $(\xi_j, \eta_j)_{\mathrm{AO}}$  (see Figure~\ref{img:intersections}). This method allows us to create a new model contour with the coordinates ($\xi_j$,~$\eta_j$)$_{\mathrm{M}}$.

While the AO silhouette remains fixed in size, the dimension of the model silhouette is parametrized by a scale $c$. The shift between the centers of the silhouettes is parametrized by an offset ($\xi_0$, $\eta_0$), which is also optimized \citep[for more details see][]{Durech2011}. In the case, we have multiple AO observations of the same asteroid taking at different epochs, we optimize one scale value for all AO and model contours. On the other hand, the offset is different for each pair of contours. We minimize the function:

\begin{equation}\label{eq:chi_ao1}
 \chi^2=\sum_{i=1}^{N}\,\sum_{j=1}^{n_i}\,\frac{[ (\xi_j, \eta_j)_{\mathrm{AO}}^{(i)} - (\xi_j, \eta_j)_{\mathrm{M}}^{(i)} ]^2}{(\sigma_j^{i})^2},
\end{equation}
where $N$ is the number of AO images, $n_i$ the number of points defining the contour of the $i$-th AO image, $(\xi_j, \eta_j)_{\mathrm{AO}}^{(i)}$ the contour of the $i$-th AO image, $(\xi_j, \eta_j)_{\mathrm{M}}^{(i)}$ the corresponding $i$-th model contour and $(\sigma_j^{i})^2$ are errors of $(\xi_j, \eta_j)_{\mathrm{AO}}^{(i)}$.

The measure (\ref{eq:chi_ao1}) can be rewritten with parameters $c$ and $(\xi_0, \eta_0)^{(i)}$ as follows:

\begin{equation}\label{eq:chi_ao2}
 \chi^2=\sum_{i=1}^{N}\,\sum_{j=1}^{n_i}\,\frac{ (\xi_{j\mathrm{AO}}^{(i)} - c\, \xi_{j\mathrm{M}}^{(i)} - \xi_0^{(i)})^2 + (\eta_{j\mathrm{AO}}^{(i)} - c\, \eta_{j\mathrm{M}}^{(i)} - \eta_0^{(i)})^2 }{(\sigma_j^{i})^2}.
\end{equation}

To find the optimal values of free parameters $c$ and $(\xi_0, \eta_0)^{(i)}$, we optimize the measure (\ref{eq:chi_ao2}) by a simplex minimization method.

\subsection{Updated non-convex shape models for a few asteroids}

\begin{figure}
	\begin{center}
	 \resizebox{0.5\hsize}{!}{\includegraphics{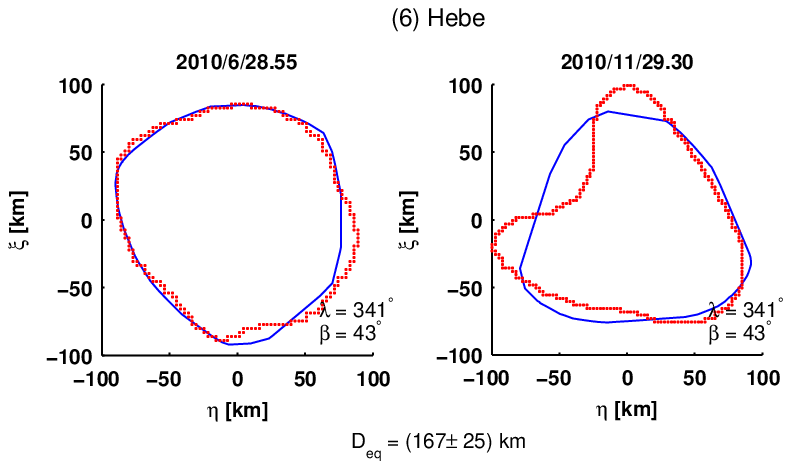}}\\
	 \resizebox{0.5\hsize}{!}{\includegraphics{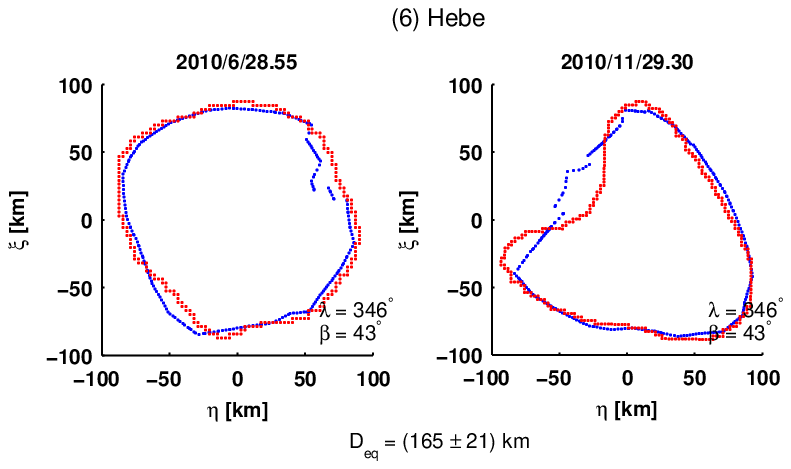}}\\
	 \end{center}
	 \caption{\label{img:6}(6) Hebe: Comparison between the AO contours (red dots) and the corresponding convex (top panel) and non-convex (bottom panel) shape model projections (blue lines).}
\end{figure}

\begin{figure}
	\begin{center}
	 \resizebox{0.5\hsize}{!}{\includegraphics{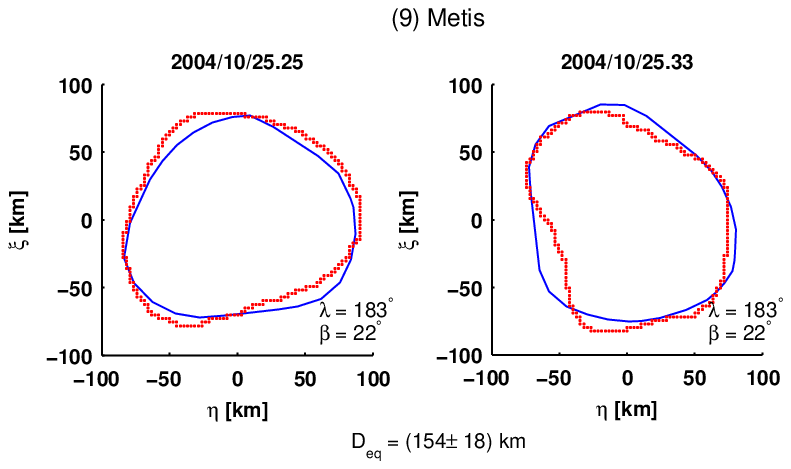}}\\
	 \resizebox{0.5\hsize}{!}{\includegraphics{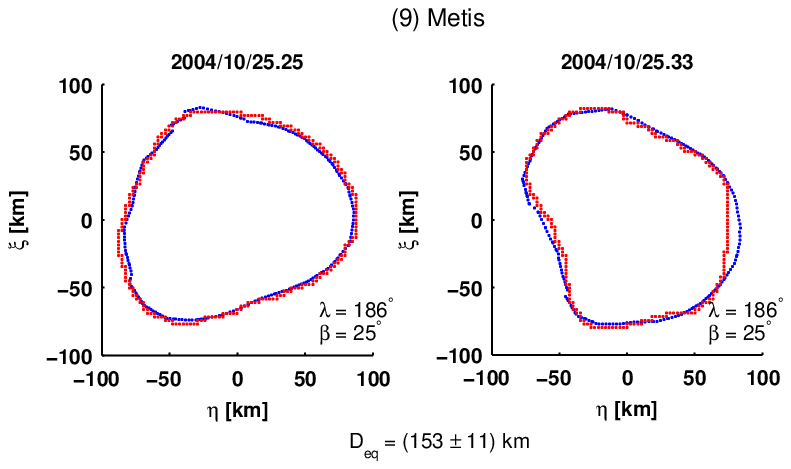}}\\
	 \end{center}
	 \caption{\label{img:9}(9) Metis: Comparison between the AO contours (red dots) and the corresponding convex (top panel) and non-convex (bottom panel) shape model projections (blue lines).}
\end{figure}

\begin{figure}
	\begin{center}
	 \resizebox{0.66\hsize}{!}{\includegraphics{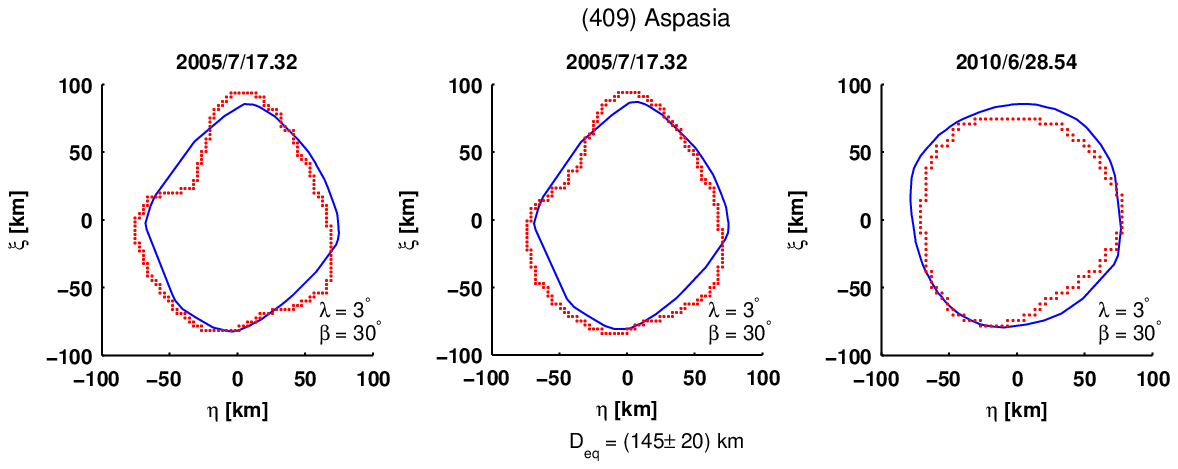}}\\
	 \resizebox{0.66\hsize}{!}{\includegraphics{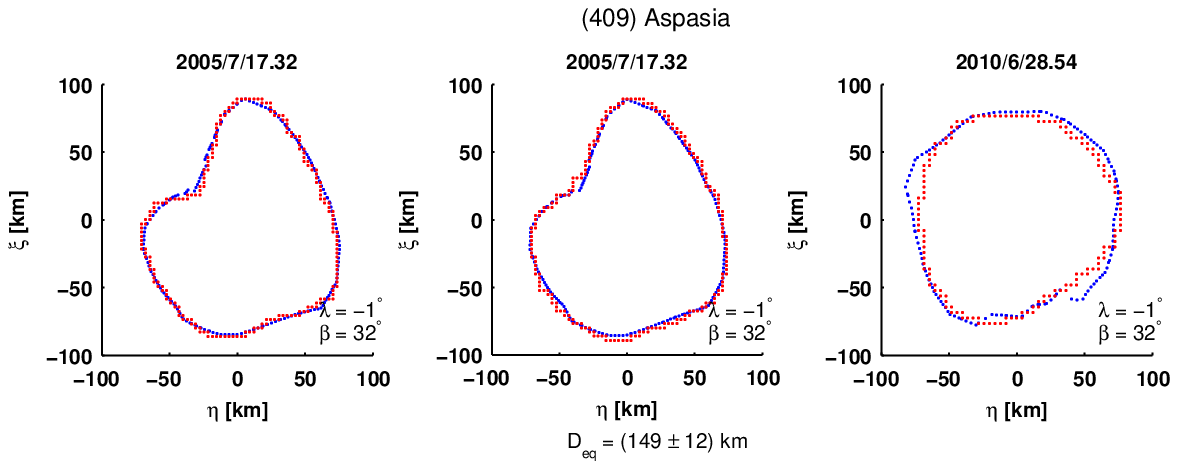}}\\
	 \end{center}
	 \caption{\label{img:409}(409) Aspasia: Comparison between the AO contours (red dots) and the corresponding convex (top panel) and non-convex (bottom panel) shape model projections (blue lines).}
\end{figure}

AO contours of asteroids (6)~Hebe, (9)~Metis and (409)~Aspasia contain non-convex features (see Figs.~\ref{img:6},~\ref{img:9}~and~\ref{img:409}), and thus comparing these contours with projections of convex shape models would lead to inaccurate sizes. To overcome this difficulty, we model these asteroids with the KOALA software (Carry et al. 2012). This multi-data set inverse technique takes into consideration, besides the disk-integrated photometry, the AO contours and converge to a usually non-convex shape solution. 

\textbf{(6) Hebe:} We failed to determine with the KOALA technique a shape solution that reproduces the non-convex feature of the AO contour (Fig.~\ref{img:6}). \citet{Torppa2003} suggested a presence of moderate albedo variegations associated with some large, flat shape features. However, these features are not situated close to position of the non-convexity. So, rather than a big area with different albedo (e.g., large crater), the concavity in the AO contour could be caused by a shadow of some terrain feature. The high phase angle of observation of $\sim$30$^{\circ}$ and the convenient position of the Sun with respect to the non-convexity feature support this scenario. This issue could be resolved by recording new AO observation closer to the opposition. With the exception of this non-convex feature, the non-convex model fits well the AO contours and so can be used for the size determination. The non-convex shape model has a pole solution (345$^{\circ}$, 42$^{\circ}$) close to the pole solution of the convex shape model (340$^{\circ}$, 42$^{\circ}$), moreover, both models have, within their uncertainties, the same rotational periods.

\textbf{(9) Metis:} The non-convex shape model complies both AO contours, and thus the size derived from this shape model is more reliable and accurate than the size determined from the convex model (Fig.~\ref{img:9}). The non-convex shape model has a pole solution (185$^{\circ}$, 25$^{\circ}$) that is only $\sim$6$^{\circ}$ apart from the pole solution of the convex shape model (180$^{\circ}$, 22$^{\circ}$), and the same rotational period.

\textbf{(403) Aspasia:} The non-convex features of the first two AO contours are explained by the KOALA non-convex model, the third contour is close to convex and also in an agreement with the shape model (Fig.~\ref{img:409}). The pole solution (359$^{\circ}$, 32$^{\circ}$) of the non-convex shape model is similar to the pole solution of the convex shape model (3$^{\circ}$, 30$^{\circ}$), as well as the rotational period.

\subsection{Comparison between AO and the shape models}

\begin{figure}
	\begin{center}
	 \resizebox{0.50\hsize}{!}{\includegraphics{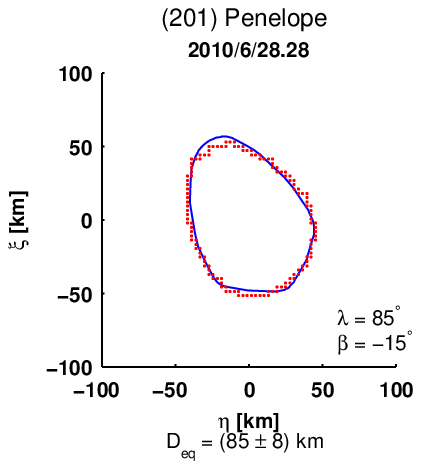}\,\includegraphics{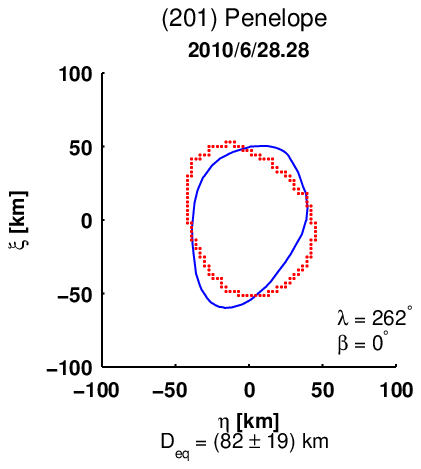}}\\
	 \end{center}
	 \caption{\label{img:201m}AO contours (red dots) and scaled silhouettes (blue lines) of two mirror solutions of asteroid (201)~Penelope. The model in the left panel is in agreement with the AO contour, on the other hand, the model in the right panel is clearly incorrect.}
\end{figure}

For the remaining 47 asteroids, we compare their AO contours with the projections of their convex shape models (all figures with the fits are included in the Supplementary material). For 42 asteroids the comparison between the AO observations and the predicted shape from our model is within the error of our models, validating independently the performance of lightcurve inversion to derive the shape of asteroids from  dense lightcurves \citep{Durech2010} and sparse-data \citep{Hanus2011, Hanus2013a}. For fifteen asteroids, we remove the pole ambiguity (see Table~\ref{tab:ao}) since the difference between the projected shape and the predicted one was obvious for one of the pole directions. Figure\ref{img:201m} illustrates the case for the asteroid (201)~Penelope.

In three cases ((22)~Kalliope, (45)~Eugenia and (165)~Loreley), the published model did not fit the contours well. After a careful analysis, we confirm that the AO contours has been properly extracted so we infer that the issue is due to the light inversion. We investigate individually each of these problematic cases.

\textbf{(22)~Kalliope:} The convex model of this asteroid was derived by \citet{Kaasalainen2002b}, but failed to fit the AO contour. The revised convex model determined by \citet{Descamps2008}, is in better agreement with the AO observation, probably due a change on the axes ratio since the pole solutions are identical within their uncertainties. We inserted the shape solution derived by \citet{Descamps2008} in the DAMIT database since it has been now validated independently.

\textbf{(45)~Eugenia:} We derive a new shape model based on additional 15 lightcurves from \citet{Marchis2010} observed during the 2007 and 2009 apparitions. By using this revised model instead of the one published by \citet{Kaasalainen2002b}, we improve the quality of the fit of the AO contours. A pole ambiguity appeared in the revised model, but was successfully removed by the AO contours. The correct pole solution agrees within the uncertainties with the one obtained by \citet{Kaasalainen2002b}, so only the shape of the asteroid has been changed. This new shape model is now part of the DAMIT database. 

\begin{figure}
	\begin{center}
	 \resizebox{0.5\hsize}{!}{\includegraphics{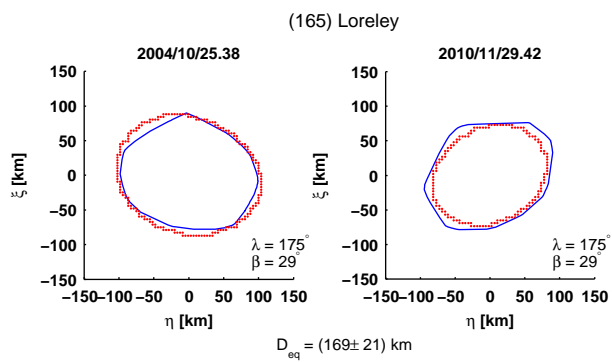}}\\
	 \end{center}
	 \caption{(165) Loreley: Comparison between the AO contours (red dots) and the corresponding convex shape model projections (blue lines).}\label{img:165}
\end{figure}

\textbf{(165)~Loreley:} The size of one of the two available contours is under- or overestimated, but it is not clear, which one is it (see Fig.~\ref{img:165}). The uncertainty of the derived size encompasses this systematic error.

Observation of asteroids \textbf{(135)~Hertha} and \textbf{(471)~Papagena} are resolved only in one direction. They agree well with the predicted orientations of the shape models. The pole ambiguity of asteroid Hertha was already removed by the occultation measurements \citep{Timerson2009}, the  AO observation confirms this pole solution. However, because their sizes cannot be estimated we discard them in the rest of the analysis. 

Observations of \textbf{(135)~Hertha} collected on September 9 2008 suggesting the binary or bilobated nature of the asteroid (see the basic AO contour in Fig.~\ref{img:135} and the basic-processed frames in the Supplementary material, Fig~\ref{img:hertha_AO}). The AO observations were recorded at a geometry close to the maximum elongation of the asteroid, so very  favorably to reveal  a possible binarity. We computed a non-convex model of asteroid (135)~Hertha using KOALA optimization scheme, constrained with the photometric data set and the deconvolved AO contour, the model delivers a single, and slightly elongated, Hertha asteroid (which is also preferred by the occultation measurements, however, the occultation was not observed close to the maximum elongation). Additional observations of the asteroid with different illuminations could help reveal the true nature of (135)~Hertha.

\begin{figure}
	\begin{center}
	 \resizebox{0.25\hsize}{!}{\includegraphics{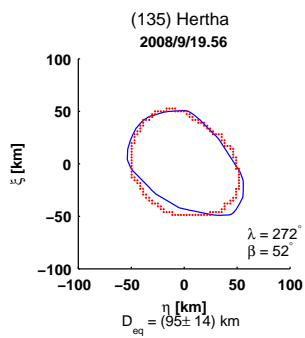}}\\
	 \end{center}
	 \caption{(135) Hertha: Comparison between the basic AO contour (red dots) and the corresponding non-convex shape model projection (blue lines).}\label{img:135}
\end{figure}

\subsection{Equivalent diameters}

Because several asteroids including (45)~Eugenia, (7)~Iris, (107)~Camilla or (130)~Elektra were observed more than once, we used all available AO observations for each asteroid simultaneously in the size optimization process described in Section~\ref{sec:ao_sizes}. The typical error in the projected major-axis dimension of the contour derived from the deconvolved image is typically 3\% \citep{Marchis2006}. The second important source of error in the size estimate comes from the convex model itself which is typically 5--8\% (1-$\sigma$ error on the fit of the contour). The overall size uncertainty is then $\sim$10\% for most of these asteroids.

For 48 asteroids ((135)~Hertha and (471)~Papagena were previously discarded from our sample of 50 asteroids, see above) with scaled convex shape models, we compute their volume using the equations from \citet{Dobrovolskis1996}, and derive the volume-equivalent diameter $D_{\mathrm{eq}}$. Table~\ref{tab:ao} contains the results of our analysis with a typical error of 10\%. For comparison, we also added the effective diameters $D_{\mathrm{IRAS}}$, $D_{\mathrm{WISE}}$, $D_{\mathrm{AKARI}}$ derived from IRAS \citep{Tedesco2002}, AKARI \citep{Usui2011} and WISE \citep{Masiero2011} and the volume-equivalent diameters $D_{\mathrm{occ}}$ derived by scaling the convex shape models to the stellar occultation measurements \citep[from][]{Durech2011}. Interestingly, asteroids for which a stellar occultation size measurement was available, have size estimate consistent with our AO-based analysis within the error bar. This is particularly encouraging since it confirms independently the reliability of our method.

\begin{figure}
	\begin{center}
	 \resizebox{\hsize}{!}{\includegraphics{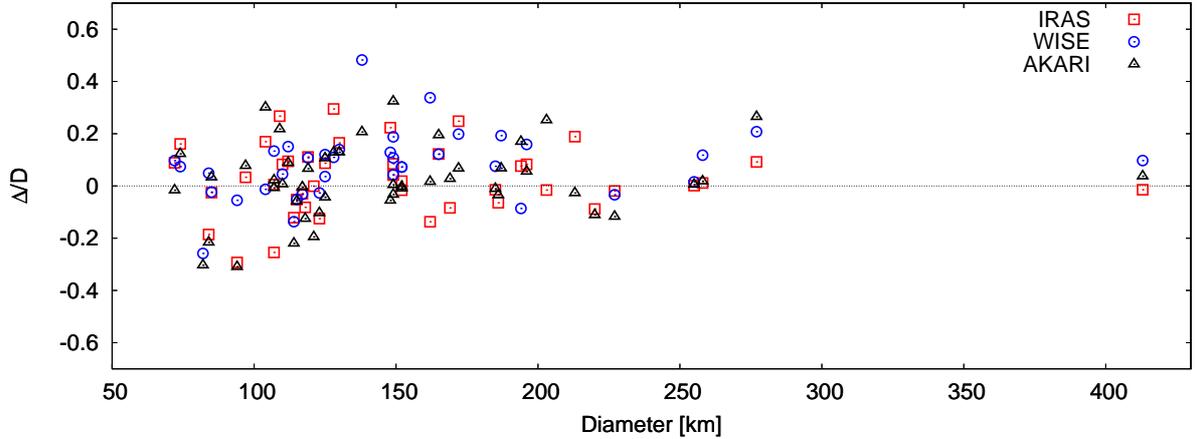}}\\
	 \end{center}
	 \caption{Relative differences between effective diameters derived by IRAS, WISE and AKARI and volume-equivalent diameters based on AO observations.}\label{img:diameter_comparison}
\end{figure}

We compute the relative differences between effective diameters derived by IRAS, WISE, and AKARI and our volume-equivalent diameters $D_{\mathrm{eq}}$ for all 48 studied asteroids and plotted them in Figure~\ref{img:diameter_comparison}. The standard deviation between the effective diameters from all three infrared surveys and volume-equivalent diameters derived here is $\sim$10\%. We notice a small systematic trend: diameters from IRAS are on average $\sim$3\% larger, diameters from WISE $\sim$7\% larger, and diameters from AKARI $\sim$2\% larger than our $D_{\mathrm{eq}}$. However, for individual asteroids, the differences in sizes (even within the infrared surveys) are often more than 20--30\%.

Because the thermal data are usually available only from one apparition, the resulting size strongly depends on the orientation of the asteroid, namely the position of its rotational axis with respect to the Earth, while a spherical shape model is usually used in the thermal modeling \citep[e.g., NEATM model of][]{Harris1998}. If the asteroid is observed pole-on, the size is overestimated (this orientation corresponds to the largest projected area of the asteroid, this area also do not change here significantly during the revolution). On the other hand, if we have thermal data observed in a configuration when the spin axis is perpendicular to the line connecting the observer and the asteroid (equator-on), the projected area is changing significantly during the revolution and could reach values between its minimum or maximum. This could result in an under- or overestimated size determination. Additionally, the observed thermal flux, which is proportional to $T^4$, and thus the derived size of the asteroid, depends on the relative geometry of the pole, the Sun, and the Earth. While we look at very different areas on the asteroid, the thermal flux could be also different, which affects the size estimate as well.

In summary, the diameters from infrared surveys are reliable in a statistical sense. For individual asteroids (especially elongated ones), the sizes could be off by up to 30\%.

%\onecolumn
\begin{landscape}
\scriptsize{
\begin{longtable}{r@{\,\,\,}l c cc ccc cc cc c cc}
\captionsetup{width=1.00\textwidth}
\caption{\label{tab:ao}List of volume-equivalent diameters $D_{\mathrm{eq}}^{(1)}$ (and $D_{\mathrm{eq}}^{(2)}$ for the mirror solution if any) for 48 scaled asteroids. The table also gives effective diameters $D_{\mathrm{IRAS}}$, $D_{\mathrm{WISE}}$, $D_{\mathrm{AKARI}}$ derived from IRAS, WISE and AKARI infrared measurements,  volume-equivalent diameters $D_{\mathrm{occ}}^{(1)}$ (and $D_{\mathrm{occ}}^{(2)}$ for the mirror solution if any, preferred solution is labeled by bold font) derived by scaling the convex shape models to the stellar occultation measurements \citep{Durech2011}, the number of AO images $N_{\mathrm{AO}}$ used for the asteroid size scaling, the mass $M$ based on \citet{Carry2012b}, the mass $M_{\mathrm{multiple}}$ determined from the moon orbits, the bulk density $\rho_{\mathrm{bulk}}$, and the Bus/DeMeo \citep[][if not available, see the table footnote for the source]{DeMeo2009} and the Tholen \citep{Tholen1984, Tholen1989} taxonomy. Pole solutions inconsistent with the AO images are marked as \textit{Rejected}.}\\
\hline 
 \multicolumn{2}{c} {Asteroid} & \multicolumn{1}{c} {$N_{\mathrm{AO}}$} & \multicolumn{1}{c} {$D_{\mathrm{eq}}^{(1)}$} & \multicolumn{1}{c} {$D_{\mathrm{eq}}^{(2)}$} & \multicolumn{1}{c} {$D_{\mathrm{IRAS}}$}  & \multicolumn{1}{c} {$D_{\mathrm{WISE}}$} & \multicolumn{1}{c} {$D_{\mathrm{AKARI}}$} & \multicolumn{1}{c} {$D_{\mathrm{occ}}^{(1)}$}  & \multicolumn{1}{c} {$D_{\mathrm{occ}}^{(2)}$} & \multicolumn{1}{c} {$M^a$} & \multicolumn{1}{c} {$M_{\mathrm{multiple}}$} & \multicolumn{1}{c} {$\rho$} & \multicolumn{1}{c} {Bus/DeMeo} & \multicolumn{1}{c} {Tholen} \\
\multicolumn{2}{l} { } &  & [km] & [km] & [km] & [km] & [km] & [km] & [km] & 10$^{18}$[kg] & 10$^{18}$[kg] & [g.cm$^{-3}$] &  & \\
\hline\hline 

\endfirsthead
\caption{continued.}\\

\hline
 \multicolumn{2}{c} {Asteroid} & \multicolumn{1}{c} {$N_{\mathrm{AO}}$} & \multicolumn{1}{c} {$D_{\mathrm{eq}}^{(1)}$} & \multicolumn{1}{c} {$D_{\mathrm{eq}}^{(2)}$} & \multicolumn{1}{c} {$D_{\mathrm{IRAS}}$}  & \multicolumn{1}{c} {$D_{\mathrm{WISE}}$} & \multicolumn{1}{c} {$D_{\mathrm{AKARI}}$} & \multicolumn{1}{c} {$D_{\mathrm{occ}}^{(1)}$}  & \multicolumn{1}{c} {$D_{\mathrm{occ}}^{(2)}$} & \multicolumn{1}{c} {$M^a$} & \multicolumn{1}{c} {$M_{\mathrm{multiple}}$} & \multicolumn{1}{c} {$\rho$} &  \multicolumn{1}{c} {Bus/DeMeo} & \multicolumn{1}{c} {Tholen} \\
\multicolumn{2}{l} { } &  & [km] & [km] & [km] & [km] & [km] & [km] & [km] & 10$^{18}$[kg] & 10$^{18}$[kg] & [g.cm$^{-3}$] &  & \\
\hline\hline 

\endhead
\hline
%\\
\multicolumn{14}{l} {$^a$\,\citet{Carry2012b}
$^b$\,\citet{Vachier2012}
$^c$\,\citet{Marchis2008}
$^d$\,\citet{Berthier2013}
$^e$\,\citet{Marchis2008a}
$^f$ SMASS II taxonomy, \citet{Bus2002}}\\
\multicolumn{14}{l} {$^g$ Taxonomic classification based on the spectra of the SMASS survey, \citet{Xu1995} 
$^h$ S3OS2, \citet{Lazzaro2004}}\endfoot

5 & Astraea & 1 & 110$\pm$14 &  & 119.1$\pm$6.5 & 115.0$\pm$9.4 & 110.8$\pm$1.4 & 115$\pm$6 &  & 2.64$\pm$0.44 &  & 3.79$\pm$1.58 & S & S \\ 
6 & Hebe & 2 & 165$\pm$21 &  & 185.2$\pm$2.9 & 185.0$\pm$10.7 & 197.2$\pm$1.8 & 180$\pm$40 &  & 13.9$\pm$1.0 &  & 5.91$\pm$1.45 & S$^f$ & S \\ 
7 & Iris & 3 & 203$\pm$24 & 202$\pm$25 & 199.8$\pm$10.0 &  & 254.2$\pm$3.3 & 198$\pm$27 & 199$\pm$26 & 12.9$\pm$2.1 &  & 2.97$\pm$1.18 & S & S \\ 
8 & Flora & 2 & 125$\pm$12 & 125$\pm$10 & 135.9$\pm$2.3 & 140.0$\pm$1.2 & 138.3$\pm$1.4 & 141$\pm$10 & \textbf{140$\pm$7} & 9.17$\pm$2.92 &  & 8.97$\pm$1.89 & Sw & S \\ 
9 & Metis & 2 & 153$\pm$11 &  &  & 204.5$\pm$3.7 & 166.5$\pm$2.1 & 169$\pm$20 &  & 8.39$\pm$1.67 &  & 4.47$\pm$1.07 & T$^h$ & S \\ 
10 & Hygiea & 1 & 413$\pm$29 & 413$\pm$20 & 407.1$\pm$6.8 & 453.2$\pm$19.2 & 428.5$\pm$6.6 & 351$\pm$27 & 443$\pm$45 & 86.3$\pm$5.2 &  & 2.34$\pm$0.34 & C & C \\
14 & Irene & 2 & 149$\pm$17 & Rejected &  & 155.4$\pm$4.4 & 144.1$\pm$1.9 &  &  & 2.91$\pm$1.88 &  & 1.68$\pm$1.23 & S & S \\ 
15 & Eunomia & 1 & 254$\pm$27 &  & 255.3$\pm$15.0 & 259.0$\pm$35.5 & 256.4$\pm$3.1 &  &  & 31.4$\pm$1.8 &  & 3.66$\pm$1.19 & K & S \\ 
16 & Psyche & 1 & 213$\pm$15 & Rejected & 253.2$\pm$4.0 &  & 207.2$\pm$3.0 & \textbf{225$\pm$20} & 225$\pm$36 & 27.2$\pm$7.5 &  & 5.38$\pm$1.87 & Xk & M \\ 
19 & Fortuna & 1 & 187$\pm$13 &  &  & 223.0$\pm$43.6 & 199.7$\pm$3.0 &  &  & 8.60$\pm$1.46 &  & 2.51$\pm$0.68 & Ch & G \\ 
22 & Kalliope & 1 & 148$\pm$17 &  & 181.0$\pm$4.6 & 167.0$\pm$15.3 & 139.8$\pm$2.1 & 143$\pm$10 &  & 7.96$\pm$0.31 & 7.75$\pm$0.70$^b$ & 4.57$\pm$1.63 & X & M \\ 
23 & Thalia & 2 & 107$\pm$12 & 107$\pm$13 & 107.5$\pm$2.2 &  & 106.2$\pm$1.9 &  &  & 1.96$\pm$0.09 &  & 3.06$\pm$1.08 & S$^f$ & S \\ 
28 & Bellona & 1 & Rejected & 121$\pm$11 & 120.9$\pm$3.4 &  & 97.4$\pm$1.4 & 97$\pm$11 & 100$\pm$10 & 2.62$\pm$0.16 &  & 2.82$\pm$0.79 & S & S \\ 
29 & Amphitrite & 1 & 196$\pm$22 &  & 212.2$\pm$6.8 & 227.1$\pm$4.0 & 206.9$\pm$2.6 &  &  & 12.9$\pm$2.0 &  & 3.27$\pm$1.21 & S & S \\ 
30 & Urania & 1 & 114$\pm$14 & 114$\pm$16 & 100.2$\pm$2.4 & 98.4$\pm$2.1 & 88.9$\pm$1.0 &  &  & 1.74$\pm$0.49 &  & 2.24$\pm$1.09 & S & S \\ 
34 & Circe & 1 & 117$\pm$14 & 116$\pm$11 & 113.5$\pm$3.3 & 113.2$\pm$2.9 & 116.5$\pm$1.1 & 96$\pm$10 & 107$\pm$10 & 3.66$\pm$0.03 &  & 4.42$\pm$1.42 & Ch & C \\ 
37 & Fides & 1 & 118$\pm$10 & Rejected & 108.3$\pm$1.9 &  & 103.2$\pm$1.4 &  &  &  &  &  & S & S \\ 
39 & Laetitia & 2 & 152$\pm$15 &  & 149.5$\pm$8.6 & 163.0$\pm$14.0 & 151.6$\pm$1.6 & 163$\pm$12 &  & 4.72$\pm$1.14 &  & 2.58$\pm$0.98 & Sqv & S \\
40 & Harmonia & 2 & 123$\pm$12 & Rejected & 107.6$\pm$6.2 & 119.7$\pm$1.3 & 110.3$\pm$1.3 &  &  &  &  &  & S & S \\ 
41 & Daphne & 1 & 186$\pm$27 &  & 174.0$\pm$11.7 &  & 179.6$\pm$2.6 & 187$\pm$20 &  & 6.31$\pm$0.11 &  & 1.87$\pm$0.82 & Ch & C \\ 
42 & Isis & 1 & 97$\pm$10 & Rejected & 100.2$\pm$3.4 &  & 104.5$\pm$1.4 &  &  & 1.58$\pm$0.52 &  & 3.31$\pm$1.49 & K & S \\ 
45 & Eugenia & 13 & 172$\pm$16 & Rejected & 214.6$\pm$4.2 & 206.1$\pm$6.2 & 183.6$\pm$2.9 &  &  & 5.79$\pm$0.14 & 5.69$\pm$0.12$^c$ & 2.14$\pm$0.60 & C$^f$ & FC \\ 
52 & Europa & 1 & 277$\pm$25 &  & 302.5$\pm$5.4 & 334.6$\pm$20.9 & 350.4$\pm$5.1 & 293$\pm$30 &  & 23.8$\pm$5.8 &  & 2.14$\pm$0.78 & C & CF \\ 
54 & Alexandra & 2 & 128$\pm$11 & Rejected & 165.8$\pm$3.4 & 142.0$\pm$14.8 & 144.5$\pm$1.8 & 135$\pm$20 & \textbf{142$\pm$9} & 6.16$\pm$3.50 &  & 5.61$\pm$3.50 & Cgh & C \\ 
68 & Leto & 1 & 112$\pm$14 & Rejected & 122.6$\pm$5.3 & 128.9$\pm$4.2 & 122.0$\pm$1.2 & 148$\pm$25 & 151$\pm$25 & 3.28$\pm$1.90 &  & 4.46$\pm$3.08 & S & S \\ 
69 & Hesperia & 1 & 109$\pm$11 & 109$\pm$11 & 138.1$\pm$4.7 &  & 132.7$\pm$1.5 &  &  & 5.86$\pm$1.18 &  & 8.64$\pm$3.14 & Xk & M \\ 
72 & Feronia & 1 & Rejected & 74$\pm$6 & 85.9$\pm$3.6 & 79.5$\pm$1.9 & 83.1$\pm$0.9 &  &  & 3.32$\pm$8.49 &  & 15.65$\pm$40.05 & STD$^g$ & TDG \\ 
%79 & Eurynome & 1 & Rejected & 112$\pm$12 & 66.5$\pm$1.6 & 72.6$\pm$1.7 & 74.8$\pm$0.9 &  &  &  &  &  & S$^f$ & S \\ 
80 & Sappho & 2 & 72$\pm$9 &  & 78.4$\pm$1.7 & 79.0$\pm$1.4 & 70.8$\pm$0.9 & 67$\pm$11 &  &  &  &  & S$^f$ & S \\ 
85 & Io & 1 & 152$\pm$15 &  & 154.8$\pm$3.8 & 163.0$\pm$18.6 & 150.7$\pm$1.9 & 163$\pm$15 &  & 2.57$\pm$1.48 &  & 1.40$\pm$0.91 & C & FC \\ 
87 & Sylvia & 2 & 258$\pm$28 &  & 260.9$\pm$13.3 & 288.4$\pm$7.6 & 262.7$\pm$3.9 &  &  & 14.8$\pm$0.0 & 14.8$\pm$1.6$^d$ & 1.65$\pm$0.56 & X & P \\ 
88 & Thisbe & 1 & 220$\pm$19 & Rejected & 200.6$\pm$5.0 &  & 195.6$\pm$2.7 & \textbf{204$\pm$14} & 220$\pm$16 & 15.3$\pm$3.1 &  & 2.74$\pm$0.90 & B$^f$ & CF \\ 
89 & Julia & 1 & 130$\pm$15 &  & 151.5$\pm$3.1 & 148.1$\pm$10.1 & 146.8$\pm$1.9 & 140$\pm$10 &  & 6.71$\pm$1.82 &  & 5.83$\pm$2.57 & K$^f$/Ld$^h$ & S \\ 
97 & Klotho & 1 & 85$\pm$9 & Rejected & 82.8$\pm$4.5 & 83.0$\pm$5.1 & 87.8$\pm$1.0 &  &  & 1.33$\pm$0.13 &  & 4.14$\pm$1.37 & Xc & M \\ 
107 & Camilla & 3 & 227$\pm$24 &  & 222.6$\pm$17.1 & 219.4$\pm$5.9 & 200.4$\pm$3.5 & 214$\pm$28 &  & 11.2$\pm$0.3 & 11.2$\pm$0.3$^c$ & 1.83$\pm$0.58 & X$^f$ & C \\ 
129 & Antigone & 1 & 124$\pm$12 &  &  & 129.5$\pm$14.8 & 119.5$\pm$1.4 & 118$\pm$19 &  & 2.65$\pm$0.89 &  & 2.59$\pm$1.15 & X & M \\ 
130 & Elektra & 3 & 185$\pm$20 &  & 182.2$\pm$11.8 & 198.9$\pm$4.1 & 183.0$\pm$2.3 & 191$\pm$14 &  & 6.60$\pm$0.40 & 6.60$\pm$0.40$^e$ & 1.99$\pm$0.66 & Ch & G \\ 
%135 & Hertha & 1 & 92$\pm$8 & Rejected & 79.2$\pm$2.0 & 77.0$\pm$7.8 & 72.8$\pm$0.9 &  &  & 1.21$\pm$0.16 &  & 2.97$\pm$0.87 & Xk & M \\ 
146 & Lucina & 2 & Rejected & 119$\pm$11 & 132.2$\pm$2.4 & 131.8$\pm$4.8 & 126.9$\pm$1.6 &  &  &  &  &  & Ch & C \\ 
152 & Atala & 1 & 82$\pm$9 &  &  & 60.8$\pm$0.9 & 57.1$\pm$1.0 & 65$\pm$8 &  & 5.43$\pm$1.24 &  & 18.81$\pm$7.54 & S$^f$ & I \\ 
165 & Loreley & 2 & 169$\pm$21 &  & 154.8$\pm$4.8 &  & 173.7$\pm$2.6 & 175$\pm$8 &  & 19.1$\pm$1.9 &  & 7.56$\pm$2.92 & C & CD \\ 
184 & Dejopeja & 1 & 93$\pm$9 & 95$\pm$9 & 66.5$\pm$2.0 & 88.8$\pm$1.1 & 64.9$\pm$0.9 &  &  &  &  &  & X$^f$ & X \\ 
201 & Penelope & 1 & 85$\pm$8 & Rejected & 68.4$\pm$3.5 & 88.1$\pm$2.8 & 65.8$\pm$1.1 &  &  &  &  &  & Xk & M \\ 
230 & Athamantis & 1 & 115$\pm$12 & 116$\pm$12 & 109.0$\pm$2.0 & 109.0$\pm$13.0 & 108.3$\pm$1.2 &  &  & 1.89$\pm$0.19 &  & 2.34$\pm$0.85 & Sl$^g$ & S \\ 
%233 & Asterope & 1 & 96$\pm$10 & Rejected & 102.8$\pm$7.9 & 99.6$\pm$2.8 & 93.0$\pm$1.0 &  &  &  &  &  & K & T \\ 
250 & Bettina & 1 & 107$\pm$15 & Rejected & 79.8$\pm$4.6 & 121.3$\pm$2.0 & 109.4$\pm$1.5 &  &  &  &  &  & Xk & M \\ 
276 & Adelheid & 1 & 104$\pm$11 & 104$\pm$12 & 121.6$\pm$7.7 & 102.7$\pm$0.7 & 135.3$\pm$2.1 & 125$\pm$15 & 117$\pm$15 &  &  &  & - & X \\ 
349 & Dembowska & 1 & Rejected & 162$\pm$17 & 139.8$\pm$4.3 & 216.7$\pm$7.4 & 164.7$\pm$1.8 &  &  & 3.58$\pm$1.03 &  & 1.61$\pm$0.69 & R & R \\ 
354 & Eleonora & 1 & 149$\pm$16 &  & 155.2$\pm$8.5 & 165.0$\pm$15.6 & 149.6$\pm$2.0 &  &  & 7.18$\pm$2.57 &  & 4.15$\pm$2.00 & A & S \\ 
409 & Aspasia & 3 & 149$\pm$12 &  & 161.6$\pm$6.8 & 177.0$\pm$0.9 & 197.2$\pm$3.7 & 173$\pm$17 &  & 11.8$\pm$2.3 &  & 6.81$\pm$1.67 & Xc & CX \\
423 & Diotima & 1 & 194$\pm$18 &  & 208.8$\pm$4.9 & 177.3$\pm$6.3 & 226.9$\pm$3.1 &  &  & 6.91$\pm$1.93 &  & 1.81$\pm$0.71 & C$^g$ & C \\ \hline
\hline

%\tablefoot{The table also gives effective diameters $D_{\mathrm{IRAS}}$, $D_{\mathrm{WISE}}$, $D_{\mathrm{AKARI}}$ derived from IRAS, WISE and AKARI infrared measurements, and volume-equivalent diameters $D_{\mathrm{occ}}^{(1)}$ (and $D_{\mathrm{occ}}^{(2)}$ for the mirror solution if any) derived by scaling the convex shape models to the stellar occultation measurements \citep{Durech2011}, the references to the convex models, and the number of AO images $N_{\mathrm{AO}}$ used for the asteroid size scaling.\\
%\footnotetext{a}{\citet{Carry2012b}}
%}
\end{longtable}
}
\end{landscape}
%\twocolumn

\section{Average densities}

To directly determine an average density $\rho$ of an asteroid, we need to know both its volume $V$ and mass $M$ ($\rho=M/V$). The volume can be computed directly from the scaled shape model derived in Section~\ref{sec:ao_sizes} using equations from \citet{Dobrovolskis1996}, or alternatively via the volume-equivalent diameters. Our main source of mass estimates was the compilation of masses by \citet{Carry2012b}. The author searched the literature for mass estimates of asteroids determined by several techniques (such as orbit deflection during a close encounter, planetary ephemeris, spacecraft tracking, or orbit of a satellite). He gathered the mass of $\sim$250 asteroids, selecting the best estimate if more than one mass value was available for an asteroid. For a few multiple asteroids with a known mutual orbit, we used the derived mass measurements from the corresponding publications since these masses are accurate to within 7\%. Ultimately, we extracted the masses of 40 asteroids from our sample (see Table~\ref{tab:ao}).

From this selected sample, we compute the bulk densities $\rho$ for 40 asteroids and include them in Table~\ref{tab:ao}. If a model is still ambiguous (i.e., with two mirror pole solutions), we derive the density for both models and list the average value in Table~\ref{tab:ao}. The uncertainties in density $\delta\rho$ are computed by the relation

\begin{equation}\label{eq:rho_error}
 \frac{\delta\rho}{\rho}=\sqrt{\left(\frac{\delta M}{M}\right)^2+\left(\frac{\delta V}{V}\right)^2},
\end{equation}

where $\delta M$ and $\delta D_{\mathrm{eq}}$ are mass and volume-equivalent diameter uncertainties, respectively. The uncertainties $\delta M$ of adopted masses usually correspond to 1-$\sigma$ level. 

While $\rho \sim M/D^3_{\mathrm{eq}}$, the uncertainty on the density should be dominated by the size uncertainty. However, the size error, which is usually $\sim$10\%, is in many cases significantly lower than the error of the mass ($\sim$20--50\%, depends on the method used for the mass determination). This contributes to the overall density error similarly to the mass uncertainty. If we assume an uncertainty in the size of 10\%, the uncertainty in mass needs to be $\sim$30\% to contribute equally to the density error.

According to \citet{Kaasalainen2012}, the volume uncertainty of a shape model is similar to the uncertainty in size when the shape uncertainty is not dominated by a scale factor for size, which is the case for shape models derived by the KOALA method. When we have a model previously derived by a convex inversion, we scale the size to fit the AO contours. The volume uncertainty is then computed by Eq.~(\ref{eq:rho_error}). When using the KOALA method, we optimize the shape and the size simultaneously, which results in a volume uncertainty that should be similar to the size uncertainty rather than being three times larger. To be sure not to underestimate the volume uncertainty for the three models derived by the KOALA method ((6)~Hebe, (9)~Metis, (409)~Aspasia), we use a value of 1.5 times the size uncertainty.

The masses for five asteroids ((14)~Irene, (52)~Europa, (54)~Alexandra, (72)~Feronia, and (85)~Io) have uncertainties higher than 50\%, and their computed densities have an even larger uncertainty. In this case these measurements are discarded in the subsequent discussion of the derived densities.

Such densities are more reliable than densities based on the sizes determined from mid-IR observations, because these photometric methods assume spherical shapes for the asteroids. As shown in Figure~\ref{img:diameter_comparison},  the sizes of individual objects could be over- or underestimated by even more than 30\% due to the observation geometry.

\begin{figure}
	\begin{center}
	 \resizebox{\hsize}{!}{\includegraphics{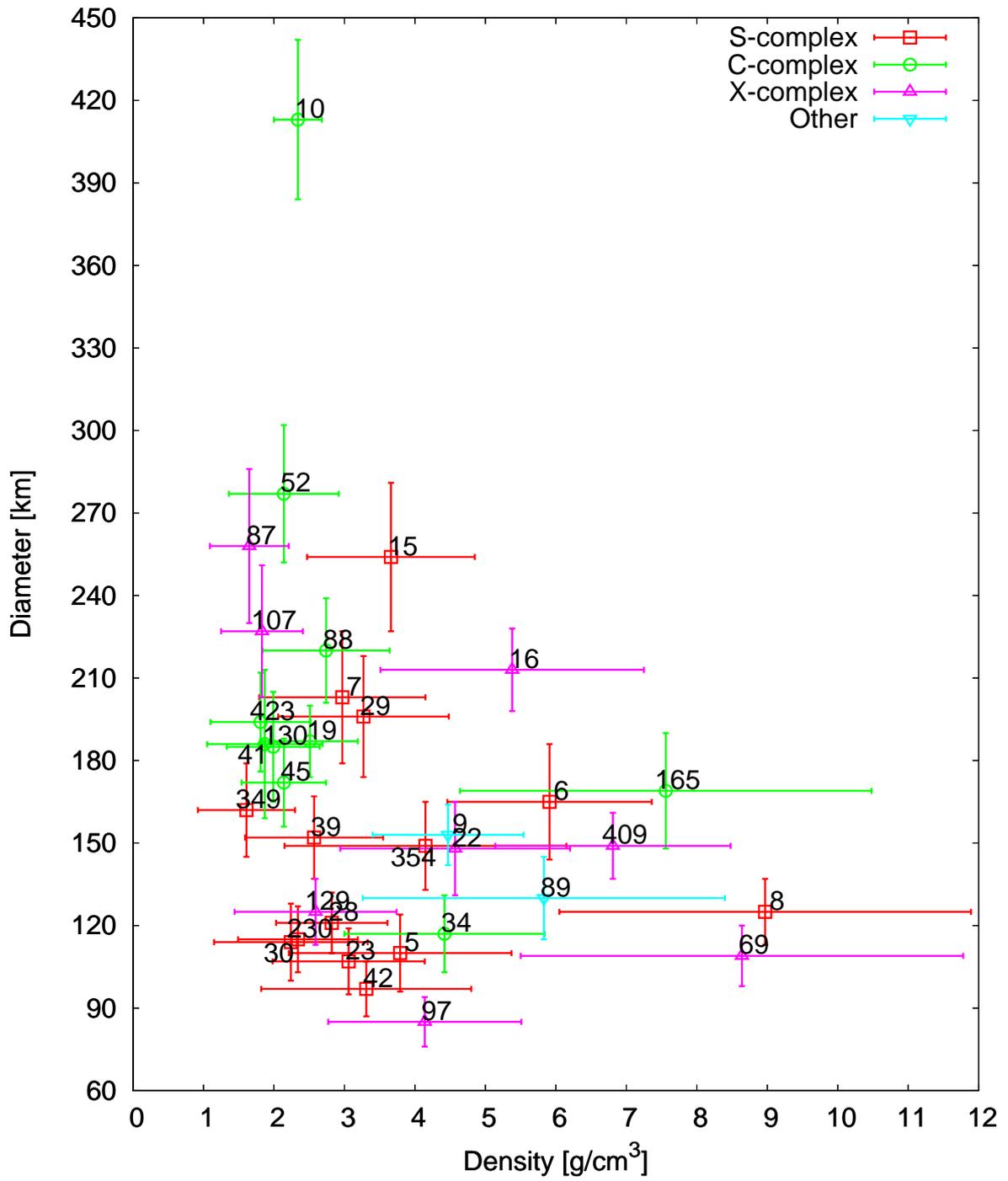}}\\
	 \end{center}
	 \caption{Dependence of asteroid equivalent diameters $D_{\mathrm{eq}}$ on their densities $\rho$ for SMASS II S, C and X taxonomic complexes and other outliner asteroid types.}\label{img:rho_vs_D}
\end{figure}

Consequently, in  Figure\ref{img:rho_vs_D}, we plot the dependence of asteroid diameters on derived bulk densities for different taxonomic complexes (S, C, X) according to the Bus/DeMeo taxonomy \citep[][see Table~\ref{tab:ao} for 35 out of 40 asteroids from our initial sample]{DeMeo2009}. If the Bus/DeMeo taxonomy was not available, we use the SMASS II, \citet{Bus2002}, SMASS, \citet{Xu1995}, or S3OS2 taxonomies, \citet{Lazzaro2004}. 

Several trends are suggested in Figure\ref{img:rho_vs_D}, even though they are preliminary due to the mass and density uncertainty of the asteroids. Large C-complex and S-complex asteroids ($D > 150$~km) have a lower bulk density, implying a larger macro-porosity than small asteroids. X-complex are more dispersed which could confirm that this group encompasses asteroids with different compositions. It is unrealistic to have asteroids with densities larger than 4 g/cm$^3$ for C-complex  and S-complex since their meteorite analogs never reach this grain density \citep{Consolmagno2008}. Similarly, based on the density of iron meteorites, metal rich asteroids, part of X-complex, could have a bulk density up to 8 g/cm$^3$. Consequently, it is very likely that the masses of the C-complex asteroid (165)~Loreley, S-complex (6)~Hebe, (8)~Flora, (152)~Atala and X-complex (69)~Hesperia are overestimated (discussed in the following subsections). New mass measurements from perturbations after close encounters measured by all-sky astrometric surveys (Gaia, Pan-STARRS, LSST) will provide more reliable values.

We include the taxonomic types C (5 asteroids), Ch (4), and B (1) into the C-complex, S (8), Sw (1), Sqv (1), Sl (1), A (1), R (1), and K (2) into the S-complex, and X (4), Xk (2), and Xc (2) into the X-complex. The remaining two asteroids cannot be directly associated with any of these three complexes, and are labeled as other and discussed in a separate section. 

\subsection{C-complex asteroids}

Ten asteroids from our sample belong to the C-complex group which is the most common type for outer asteroids in the main belt. Eight out of ten of these asteroids have a density between 1.0 and 2.7 g/cm$^3$ with an average value of 2.19 g/cm$^3$. There is no obvious correlation between the size of these 8 asteroids and their average density. Two C-complex asteroids (34)~Circe and (165)~Loreley have a density significantly different from the other members of this group with a density of 4.4$\pm$1.4 g/cm$^3$ and 8$\pm$3 g/cm$^3$, respectively. The density of (165)~Loreley is suspiciously high. Our size measurement for this asteroid (169$\pm$21 km) is in agreement with radiometric size measurements from IRAS and AKARI data. It is also very close to the size derived from occultation data (175$\pm$8 km). \citet{Carry2012b} classified this density in the unrealistic category and our work confirms that this could be due to an overestimation of its mass by a factor of $\sim$3.5. The case for (34)~Circe is similar. Our size measurement is within the error of the radiometric measurements and occultation data (see Table~\ref{tab:ao}). However, the mass is most likely overestimated by a factor of $\sim$2 leading to an unrealistic bulk density for a C-type asteroid.

The B-type asteroid (88)~Thisbe has a density of 2.74$\pm$0.90 g/cm$^3$. This high density is in agreement with the density estimates made by \citet{Carry2012b}, who showed that B-type asteroids have larger average densities ($\sim$2.4 g/cm$^3$) than other types in the C-complex.

\subsection{S-complex asteroids}

Because the most common asteroids in the inner main-belt have a high albedo, and thus are bright enough for the Keck AO system, our sample of asteroids suffers from a selection effect. Consequently, at least fifteen of them (44\%) are classified as members of the S-complex. Their average density estimates, with the exception of three of them, are between 1.6 and 4.5 g/cm$^3$ with an average of 2.98 g/cm$^3$. The densities of asteroids (6)~Hebe, (8)~Flora and (152)~Atala are larger than $\sim$6 g/cm$^3$, which is unrealistic. The relative differences between radiometric size measurements and sizes derived from occultation data and here derived equivalent diameters are $\sim$15\% for Hebe, $\sim$10\% for Flora, and  $\sim$-30\% for Atala. \citet{Carry2012b} classified densities for (8)~Flora and (152)~Atala in the unrealistic category and our work suggests that this could be due to an overestimation of their masses by a factor of $\sim$3 and $\sim$6, respectively. The density of (6)~Hebe is not dramatically high (5.91$\pm$1.45 g/cm$^3$), considering its significant uncertainty, we could get a value that is on the high end of the reasonable densities for S-complex asteroids. We can get a realistic density for Hebe also by assuming a mass overestimation by a factor of $\sim$1.5-2.

\subsection{X-complex asteroids}

Eight asteroids from our sample belong to the X-complex group. Their bulk densities vary from 1.8 to 8.6 g/cm$^3$ with an average value of 4.45 g/cm$^3$. The large spread of densities indicates a wide range of different compositions among X-complex, as already suggest by combining their reflectance spectra with albedo measurements \citep{Tholen1989b}. Some of these X-type asteroids have a composition saturated in iron-nickel metal (e.g., (22)~Kalliope), others seem to have density close to C-complex asteroids (e.g., (87)~Sylvia or (107)~Camilla).

\subsection{Outlying spectral class asteroids}

(9)~Metis and (89)~Julia have visible spectrum characteristics that lie outside the ranges of C-, X- and S-complexes. (9)~Metis is classified as a T-type asteroid by \citet{Lazzaro2004} and S-type by \citep{Tholen1989}. Its density (4.5$\pm$1.1 g/cm$^3$) is in the range, but upper limit, of S-complex asteroid, so could have a similar composition.

The case of (89)~Julia is unclear since using the same taxonomic class \citet{Bus2002} found out that its visible reflectance spectrum makes it part of the K-type (an end-member class of the S-complex), whereas \citet{Lazzaro2004} placed it in the Ld-type after having collected new data. With a bulk density of 5.8 g/cm$^3$ and a large 1-$\sigma$ error of 2.6 g/cm$^3$, it is not possible to assess its composition. We can merely notice that this bulk density is in agreement with the one derived for (15)~Eunomia, another K-type asteroid with a density of 3.3$\pm$1.5 g/cm$^3$.

\section{Conclusion}

In this work, we derive the volume-equivalent diameters for 48 asteroids with typical uncertainties lower than 10\%, caused by both the uncertainty in the size of the AO contour and the convex shape model imperfections (Table~\ref{tab:ao}) and remove the pole ambiguity of 15 asteroid models.

The asteroids (135)~Hertha and (471)~Papagena were resolved only in one direction and thus are not used for the size determination.

For 3 out of 48 studied asteroids, we notice a significant difference between our AO contour and the silhouette from the lightcurve inversion shape model. We investigate these cases. We create revised convex shape models of asteroids (22)~Kalliope and (45)~Eugenia by using additional photometric data. Although rotational state solutions are in both cases similar within their uncertainties to that of the original models, we obtain a significantly better agreement between the contours of the convex models and AO observations. We know that the size of one of the two AO contours of asteroid (165)~Loreley is affected by a systematic error, but we have been unable to distinguish which one. The error of the derived size encompasses this issue.

For two asteroids, we had to revise their shape models (asteroids (22)~Kalliope and (45)~Eugenia) to obtain a good agreement with the AO observations. This means that some of the already lightcurve-inversion shape models do not well reflect the asteroid real shape appearance, and while we investigated here only 50  asteroids from $\sim$300 for which a shape model was published, there probably exist other shape models with similar problems. However, the rotational state (i.e., the sidereal rotational period and the orientation of the spin axis) is most likely properly determined in those cases and the difference is rather in the shape itself. We conclude that regular revisions of already published models is important and should be done every time new photometric data are available. This task will be doable in a near future thanks to  the huge amount of calibrated sparse-in-time photometric observations from projects such as Pan-STARRS \citep{Kaiser2010}. Rather than a static database of asteroid shape models (such as DAMIT) a dynamic one will be necessary. Each time a significant number of new observations will be available (e.g, new dense lightcurve(s), several Pan-STARRS observations) the revised model should be computed and included in the database. Additionally, larger number of photometric observations could yield to higher shape resolution and to lower uncertainties in the rotational state.

The good agreement between AO contours and convex shape model projections for majority of studied asteroids shows that the lightcurve inversion technique provides reliable shape models and thus is very well validated.

All scaled shape models with derived sizes will be uploaded to DAMIT database. The two revised solutions will completely replace the previous models. So far, several other shape models have been updated, so one should always check the most recently available models in DAMIT before using those.

For asteroids (6)~Hebe, (9)~Metis and (409)~Aspasia, their convex shape models are not able to reproduce the non-convex features in the AO contours. We model these three asteroids by the KOALA technique (we use original photometric data and the AO contours) and derive their non-convex shape models. We get a good agreement with the AO contours for asteroids (9)~Metis and (409)~Aspasia and derive their volume-equivalent diameters. The non-convexity in the AO contour of asteroid (6) Hebe is inconsistent with the KOALA model and is probably caused by a surface feature that creates a shadow (the phase angle is $\sim$30$^{\circ}$).  

We show that diameters derived from thermal observations and by scaling convex models to fit the AO images are on average consistent. Unlike the sizes based on infrared measurements, the sizes derived by comparing convex models with AO observations are not biased by the observing geometry, and thus are more reliable \citep[infrared surveys assume for fitting the thermal measurements a spherical shape model, e.g., NEATM model of][]{Harris1998}.

We adopt mass estimates for 40 asteroids and determine their bulk densities, in 35 cases with an uncertainty lower than 50\%. We discuss the density values in the C-, S- and X-complex taxonomic groups. We show that inconsistent density measurements of several asteroids could arise from their  overestimated masses.  

For the purpose of this work, we extract a sample of asteroids from our set of 250 Keck AO observations of 164 asteroids. However, only for a third of these asteroids, a convex model is available, and thus only those are analyzed in our work. Current photometric measurements are insufficient for the remaining asteroids to derive their shape models. While this concerns mainly larger (due to the resolution limits of the AO observations), and thus brighter, asteroids, dense photometric data from several apparitions are usually available for them. In many cases, only a few additional photometric observations could allow us to derive their convex models, and subsequently use these models with the AO observations to scale the sizes. So, observing such asteroids could lead to new size estimates. A different approach is to identify asteroids with the best AO measurements and try to get new photometric data for them. In both cases, the role of observers with small and intermediate telescopes is essential.

%\citet{Carry2012,Descamps2008,Kaasalainen1992b,Marchis2010,Marchis2012a,Marchis2013a,Timerson2009,vanDam2004}
%\citet{Michalowski1995,Marciniak2007,Schober1983,Usui2011,Bus2002,Tholen1984,Tholen1989,Kaiser2010,Kaasalainen2012, Lazzaro2004, Xu1995, DeMeo2009, Tholen1989b, Kaasalainen2011,Consolmagno2008}

\section*{Acknowledgements}
The work of JH and JD has been supported by grants GACR P209/10/0537 and P209/12/0229 of the Czech Science Foundation, and by the Research Program MSM0021620860 of the Czech Ministry of Education. The work of FM has been supported by the NASA grant NNX11AD62G. The observatory was made possible by the generous financial support of the W.M. Keck Foundation. The authors extend special thanks to those of Hawaiian ancestry on whose sacred mountain we are privileged to be guests. Without their generous hospitality, none of the observations presented would have been possible.

\bibliography{mybib}
\bibliographystyle{model2-names}

\newpage

\section*{Supplementary material}

\begin{figure}[!h]
	\begin{center}
	 \resizebox{1.0\hsize}{!}{\includegraphics{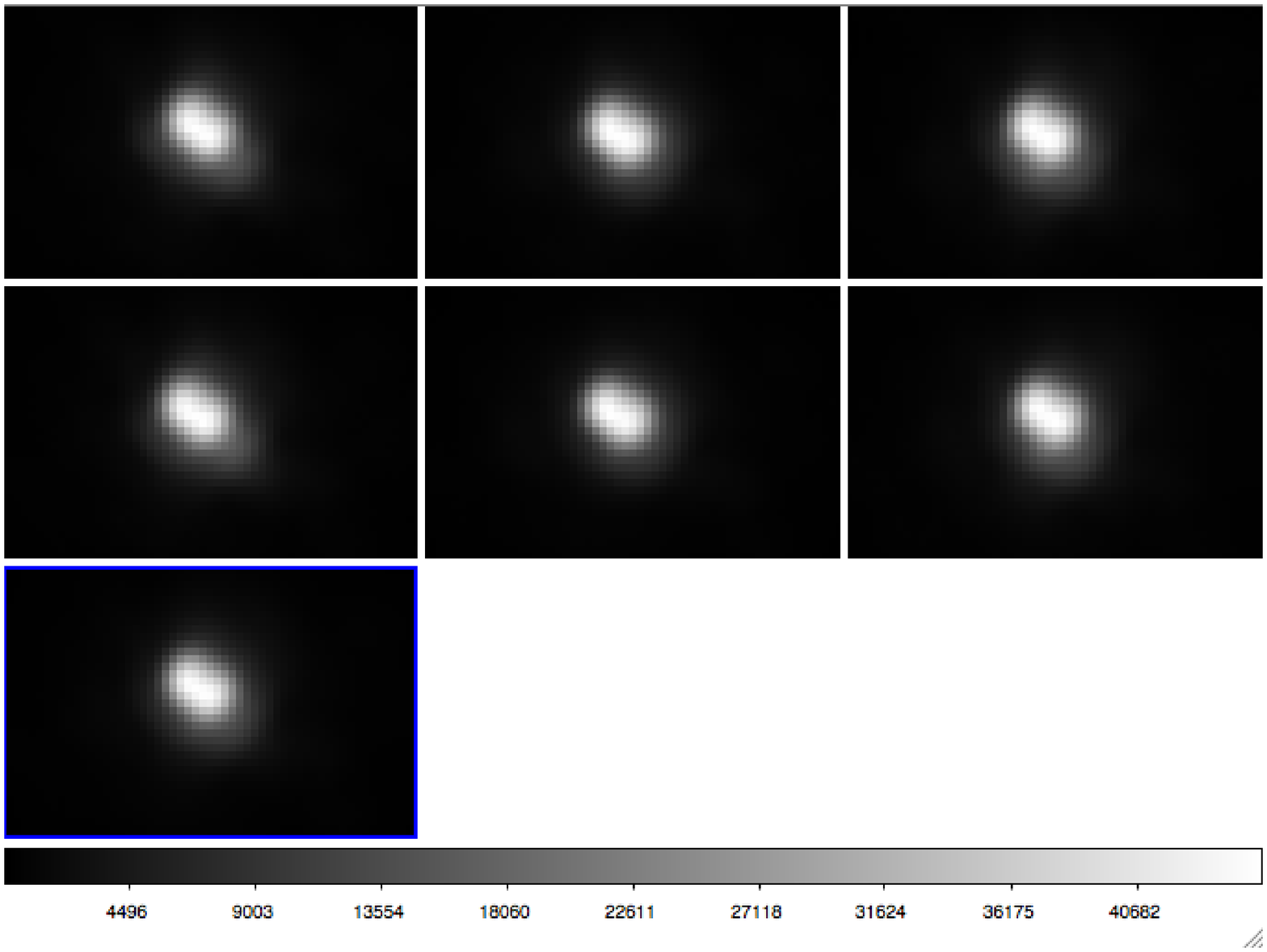}\,\includegraphics{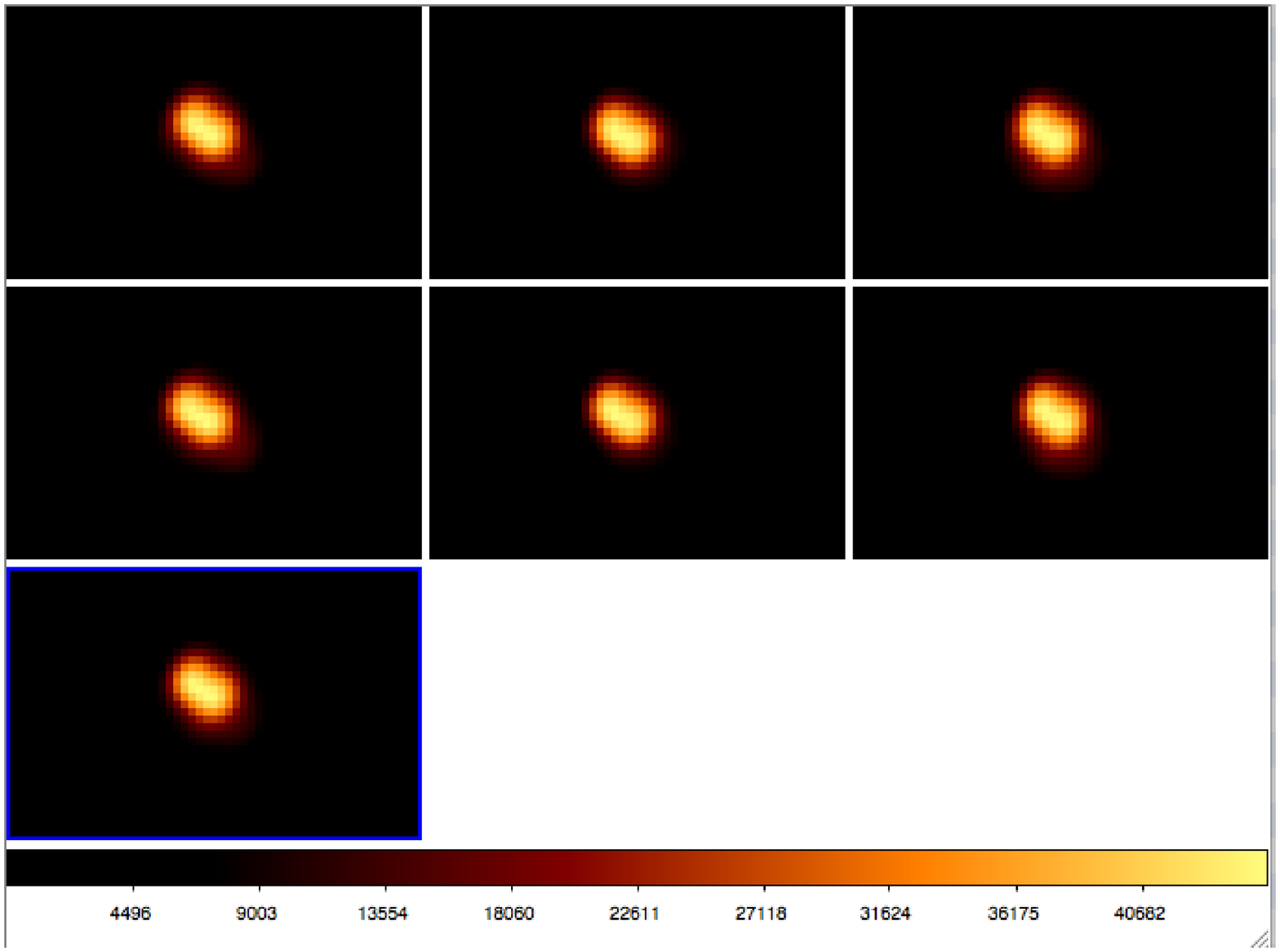}}\\
	 \end{center}
	 \caption{\label{img:hertha_AO}Observations of (135) Hertha collected on September 9 2008 suggesting the binary or bilobated nature of the asteroid. The frames shown on the first two rows correspond to the individual observations (60 s exposure time, FeII filter). The frame on the third row is the resulting shift-and-add frame. The asteroid which was observed at its maximum elongation is resolved with an angular size of 95 mas and 72 mas in its major and minor axis directions.}
\end{figure}

\newpage

\begin{figure}[!h]
	\begin{center}
	 \resizebox{0.25\hsize}{!}{\includegraphics{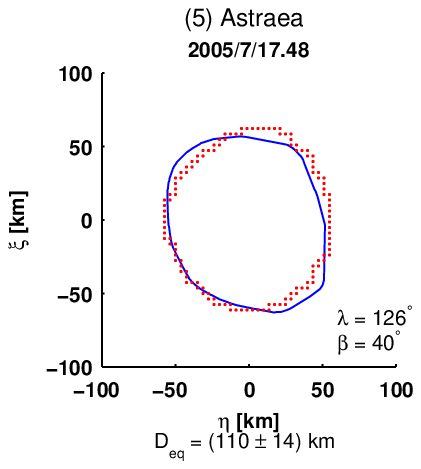}}\\
	 \end{center}
	 \caption{\label{img:5}(5) Astraea: Comparison between the AO contour (red dots) and the corresponding convex shape model projection (blue line).}
\end{figure}

\newpage

\begin{figure}[!h]
	\begin{center}
	 \resizebox{0.75\hsize}{!}{\includegraphics{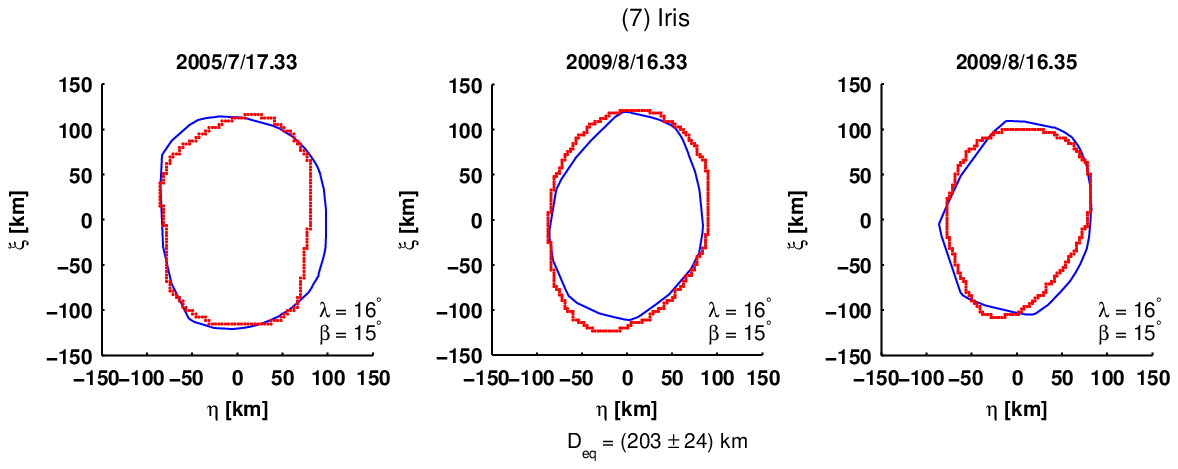}}\\
	 \resizebox{0.75\hsize}{!}{\includegraphics{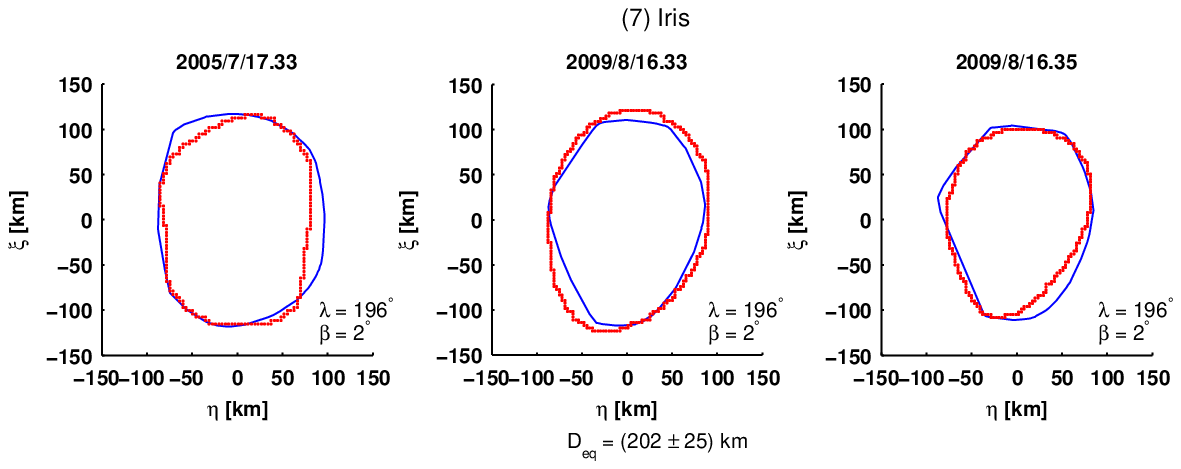}}\\
	 \end{center}
	 \caption{\label{img:7}(7) Iris: Comparison between the AO contours (red dots) and the corresponding convex shape model projections (blue line).}
\end{figure}

\newpage

\begin{figure}[!h]
	\begin{center}
	 \resizebox{1.0\hsize}{!}{\includegraphics{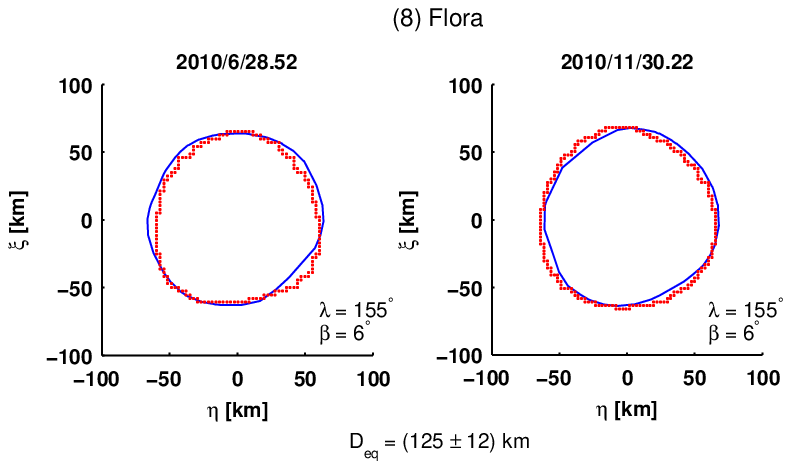}\includegraphics{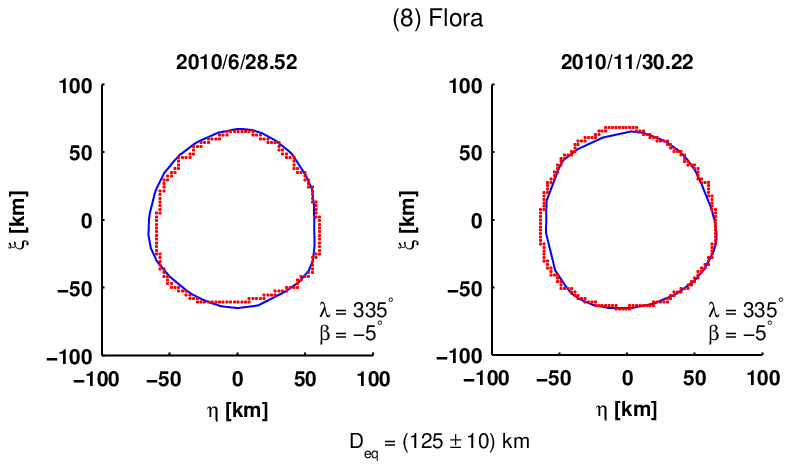}}\\
	 \end{center}
	 \caption{\label{img:8}(8) Flora: Comparison between the AO contours (red dots) and the corresponding convex shape model projections (blue line) for both pole solutions.}
\end{figure}

\newpage

\begin{figure}[!h]
	\begin{center}
	 \resizebox{0.50\hsize}{!}{\includegraphics{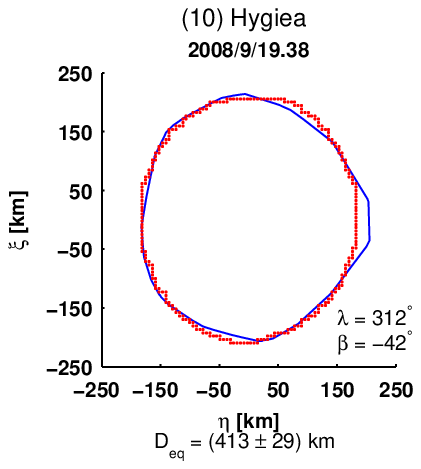}\includegraphics{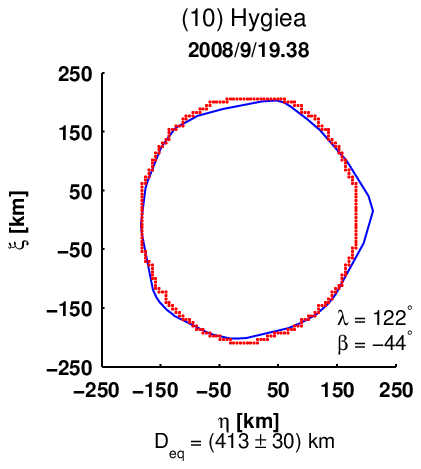}}\\
	 \end{center}
	 \caption{\label{img:10}(10) Hygiea: Comparison between the AO contour (red dots) and the corresponding convex shape model projection (blue line) for both pole solutions.}
\end{figure}

\newpage

\begin{figure}[!h]
	\begin{center}
	 \resizebox{1.0\hsize}{!}{\includegraphics{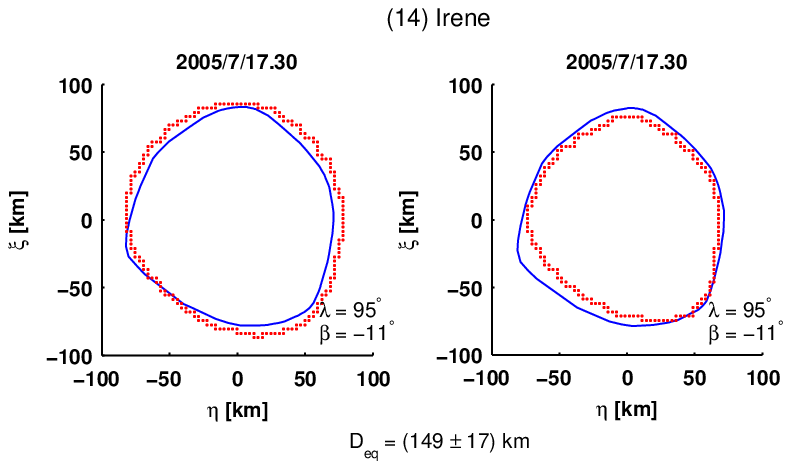}\includegraphics{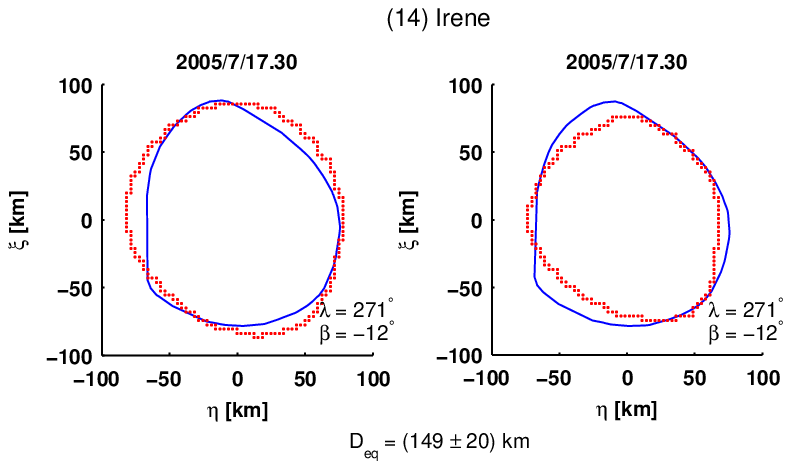}}\\
	 \end{center}
	 \caption{\label{img:14}(14) Irene: Comparison between the AO contours (red dots) and the corresponding convex shape model projections (blue line) for both pole solutions. The first pole solution is preferred.}
\end{figure}

\newpage

\begin{figure}[!h]
	\begin{center}
	 \resizebox{0.25\hsize}{!}{\includegraphics{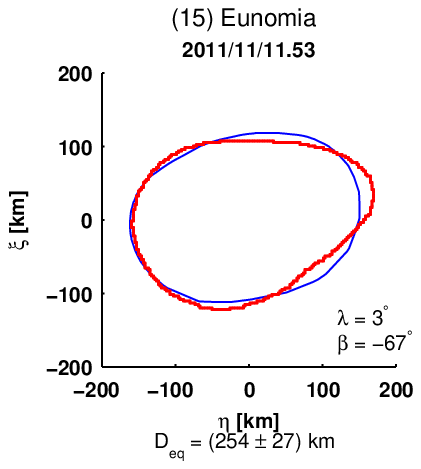}}\\
	 \end{center}
	 \caption{\label{img:15}(15) Eunomia: Comparison between the AO contour (red dots) and the corresponding convex shape model projection (blue line).}
\end{figure}

\newpage

\begin{figure}[!h]
	\begin{center}
	 \resizebox{0.50\hsize}{!}{\includegraphics{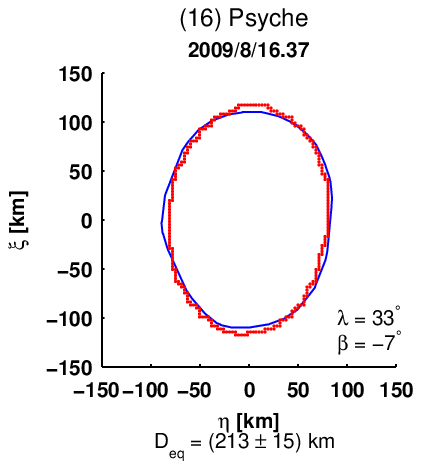}\includegraphics{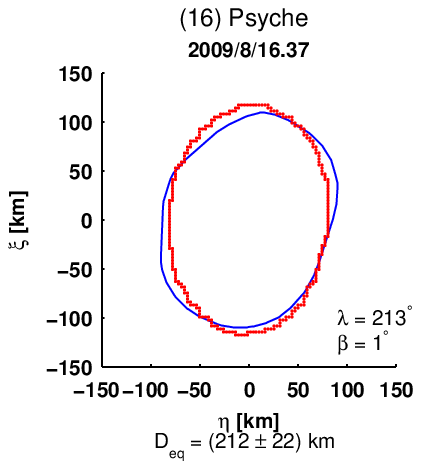}}\\
	 \end{center}
	 \caption{\label{img:16}(16) Psyche: Comparison between the AO contour (red dots) and the corresponding convex shape model projection (blue line) for both pole solutions. The first pole solution is preferred.}
\end{figure}

\newpage

\begin{figure}[!h]
	\begin{center}
	 \resizebox{0.25\hsize}{!}{\includegraphics{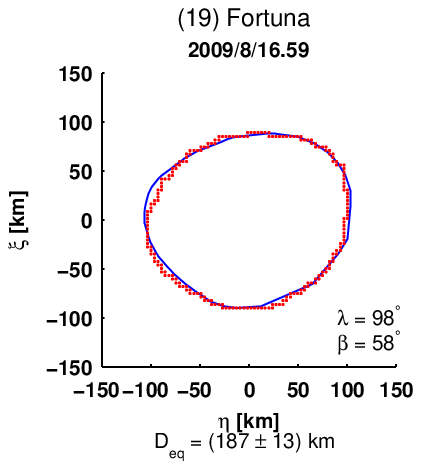}}\\
	 \end{center}
	 \caption{\label{img:19}(19) Fortuna: Comparison between the AO contour (red dots) and the corresponding convex shape model projection (blue line).}
\end{figure}

\newpage

\begin{figure}[!h]
	\begin{center}
	 \resizebox{0.25\hsize}{!}{\includegraphics{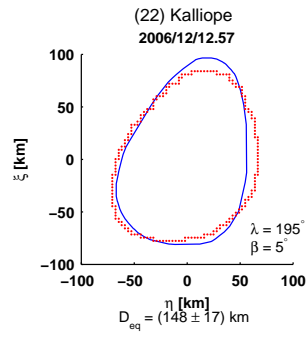}}\\
	 \end{center}
	 \caption{\label{img:22}(22) Kalliope: Comparison between the AO contour (red dots) and the corresponding projection of the convex shape model from Descamps et al (2008) (blue line).}
\end{figure}

\newpage

\begin{figure}[!h]
	\begin{center}
	 \resizebox{1.0\hsize}{!}{\includegraphics{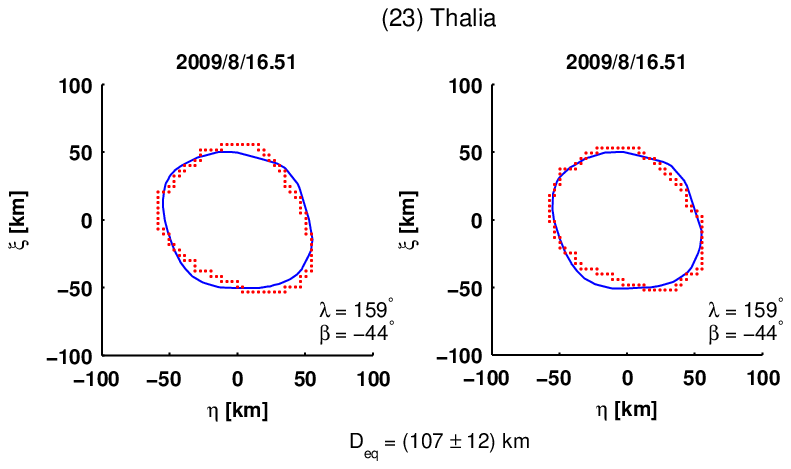}\includegraphics{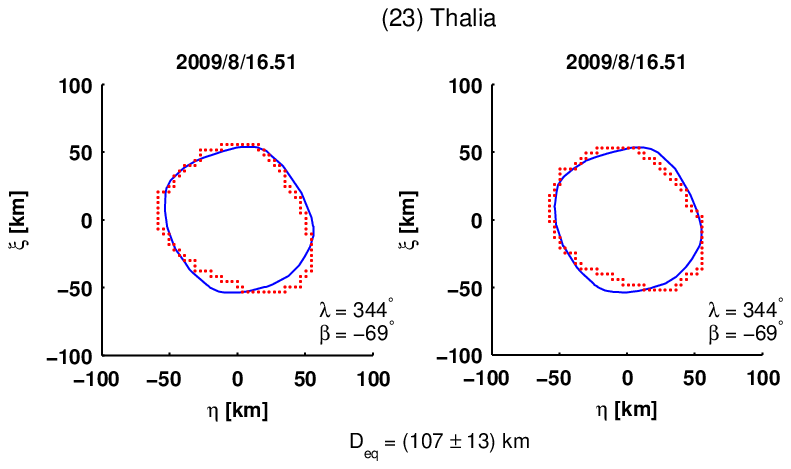}}\\
	 \end{center}
	 \caption{\label{img:23}(23) Thalia: Comparison between the AO contours (red dots) and the corresponding convex shape model projections (blue line) for both pole solutions.}
\end{figure}

\newpage

\begin{figure}[!h]
	\begin{center}
	 \resizebox{0.50\hsize}{!}{\includegraphics{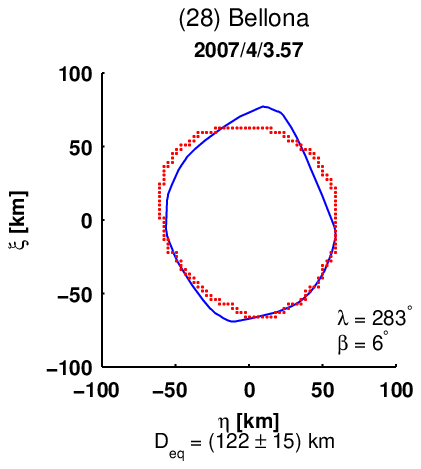}\includegraphics{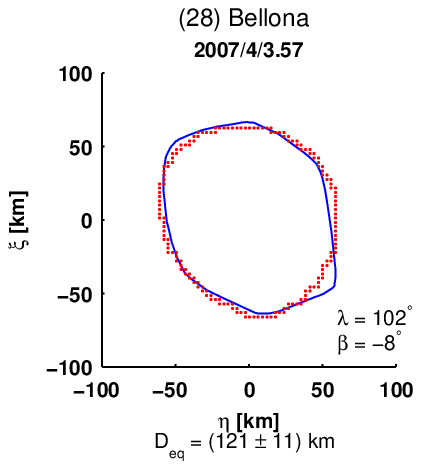}}\\
	 \end{center}
	 \caption{\label{img:28}(28) Belona: Comparison between the AO contour (red dots) and the corresponding convex shape model projection (blue line) for both pole solutions. The second pole solution is preferred.}
\end{figure}

\newpage

\begin{figure}[!h]
	\begin{center}
	 \resizebox{0.25\hsize}{!}{\includegraphics{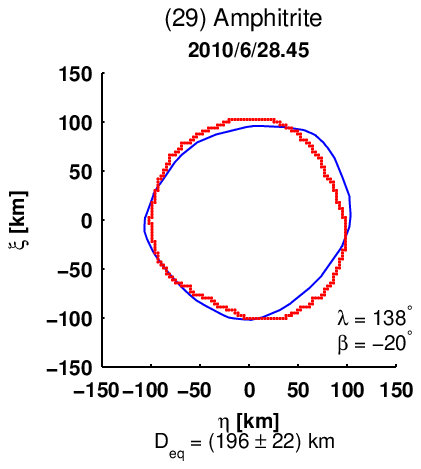}}\\
	 \end{center}
	 \caption{\label{img:29}(29) Amphitrite: Comparison between the AO contour (red dots) and the corresponding convex shape model projection (blue line).}
\end{figure}

\newpage

\begin{figure}[!h]
	\begin{center}
	 \resizebox{0.50\hsize}{!}{\includegraphics{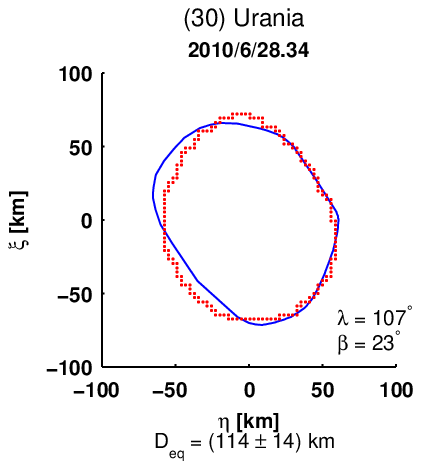}\includegraphics{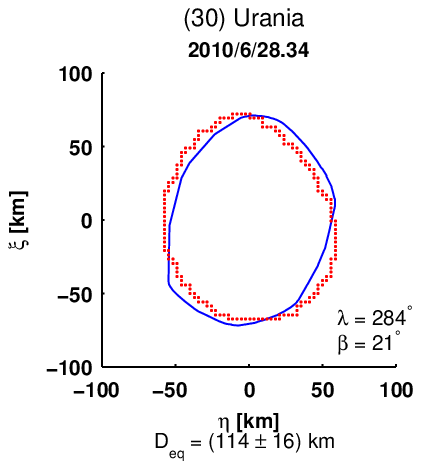}}\\
	 \end{center}
	 \caption{\label{img:30}(30) Urania: Comparison between the AO contour (red dots) and the corresponding convex shape model projection (blue line) for both pole solutions.}
\end{figure}

\newpage

\begin{figure}[!h]
	\begin{center}
	 \resizebox{0.50\hsize}{!}{\includegraphics{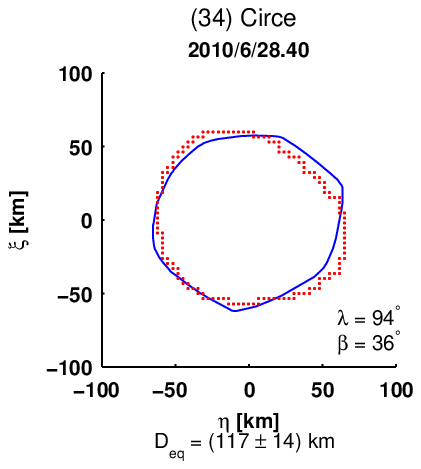}\includegraphics{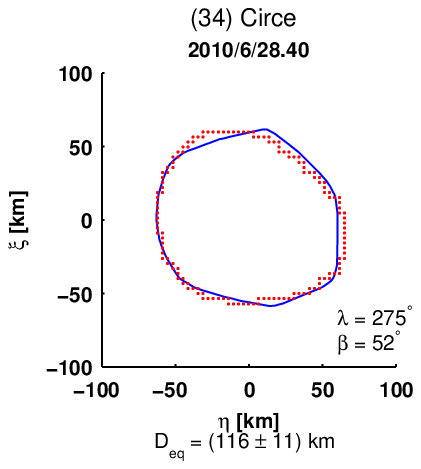}}\\
	 \end{center}
	 \caption{\label{img:34}(34) Circe: Comparison between the AO contour (red dots) and the corresponding convex shape model projection (blue line) for both pole solutions.}
\end{figure}

\newpage

\begin{figure}[!h]
	\begin{center}
	 \resizebox{0.50\hsize}{!}{\includegraphics{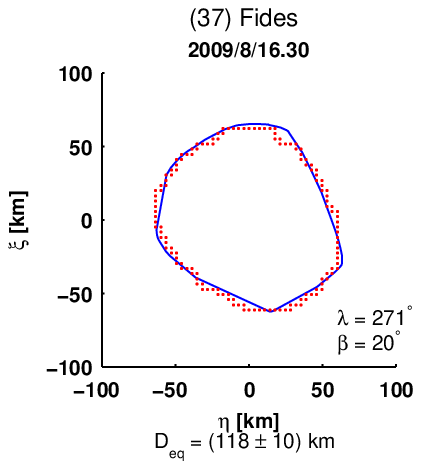}\includegraphics{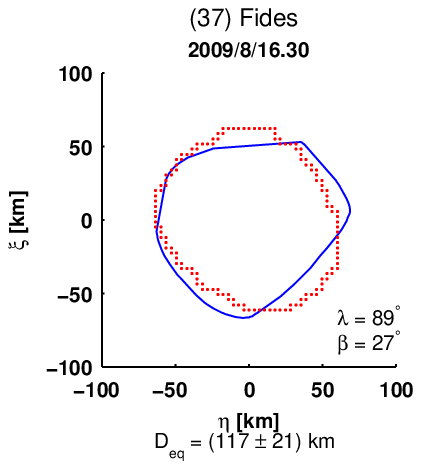}}\\
	 \end{center}
	 \caption{\label{img:37}(37) Fides: Comparison between the AO contour (red dots) and the corresponding convex shape model projection (blue line) for both pole solutions. The first pole solution is preferred.}
\end{figure}

\newpage

\begin{figure}[!h]
	\begin{center}
	 \resizebox{0.5\hsize}{!}{\includegraphics{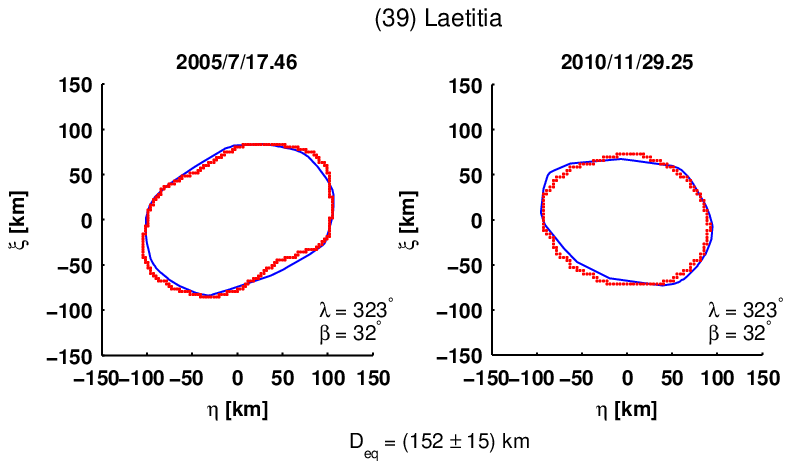}}\\
	 \end{center}
	 \caption{\label{img:39}(39) Laetitia: Comparison between the AO contours (red dots) and the corresponding convex shape model projections (blue line).}
\end{figure}

\newpage

\begin{figure}[!h]
	\begin{center}
	 \resizebox{1.0\hsize}{!}{\includegraphics{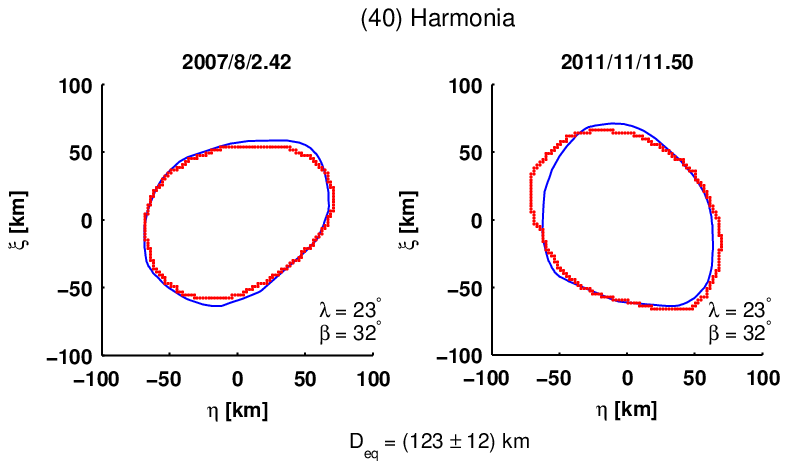}\includegraphics{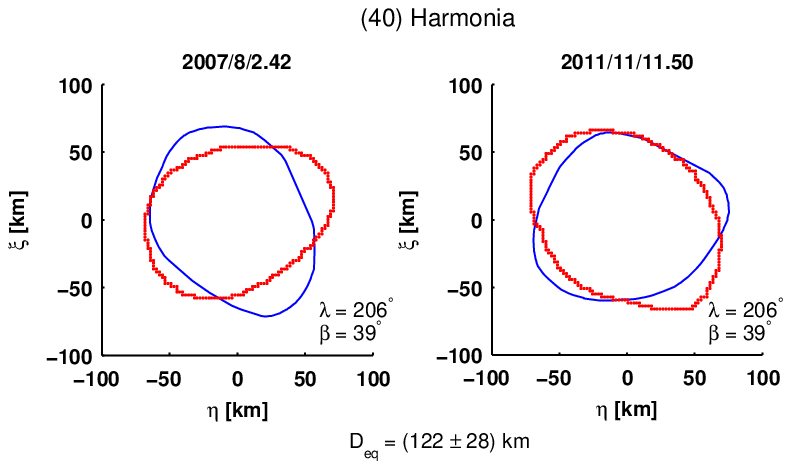}}\\
	 \end{center}
	 \caption{\label{img:40}(40) Harmonia: Comparison between the AO contours (red dots) and the corresponding convex shape model projections (blue line) for both pole solutions. The first pole solution is preferred.}
\end{figure}

\newpage

\begin{figure}[!h]
	\begin{center}
	 \resizebox{0.25\hsize}{!}{\includegraphics{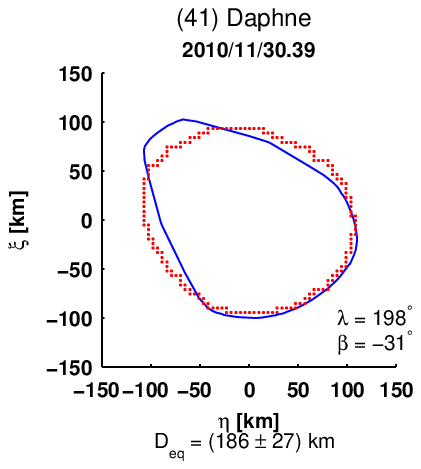}}\\
	 \end{center}
	 \caption{\label{img:41}(41) Daphne: Comparison between the AO contour (red dots) and the corresponding convex shape model projection (blue line).}
\end{figure}

\newpage

\begin{figure}[!h]
	\begin{center}
	 \resizebox{0.50\hsize}{!}{\includegraphics{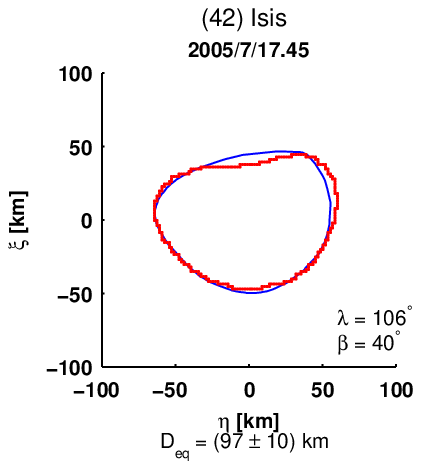}\includegraphics{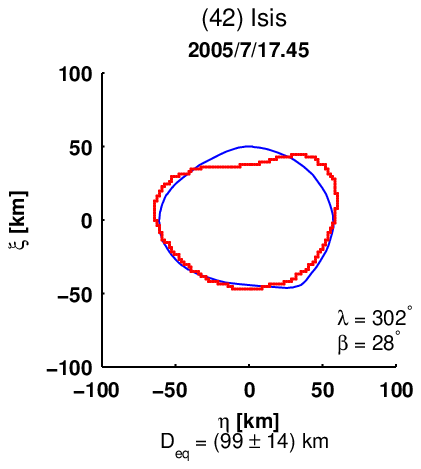}}\\
	 \end{center}
	 \caption{\label{img:42}(42) Isis: Comparison between the AO contour (red dots) and the corresponding convex shape model projection (blue line) for both pole solutions. The first pole solution is preferred.}
\end{figure}

\newpage

\begin{figure}[!h]
	\begin{center}
	 \resizebox{1.0\hsize}{!}{\includegraphics{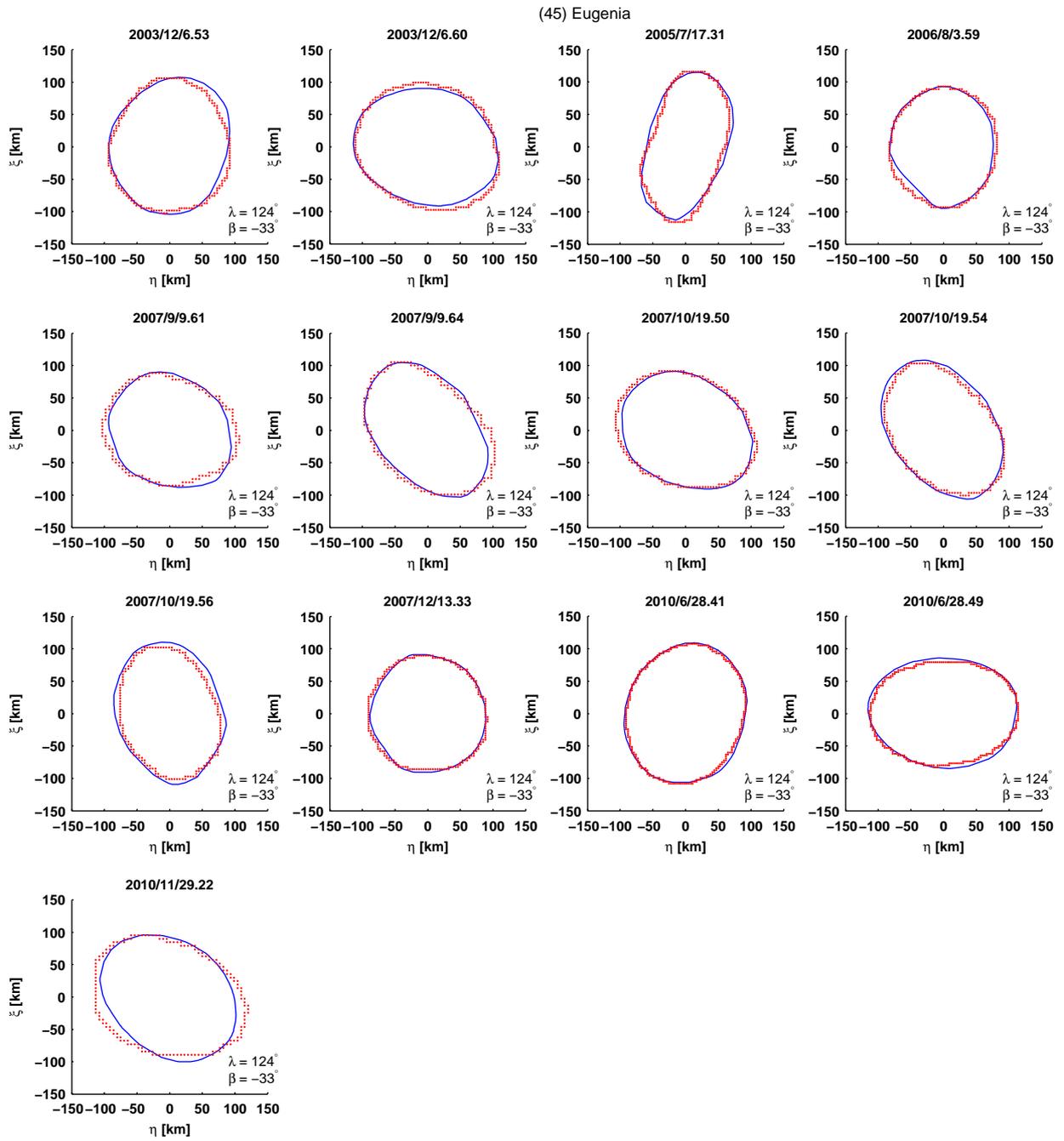}}\\
	 \end{center}
	 \caption{\label{img:45a}(45) Eugenia: Comparison between the AO contours (red dots) and the corresponding convex shape model projections (blue line) of the first pole solution, which is preferred.}
\end{figure}

\newpage

\begin{figure}[!h]
	\begin{center}
	 \resizebox{1.0\hsize}{!}{\includegraphics{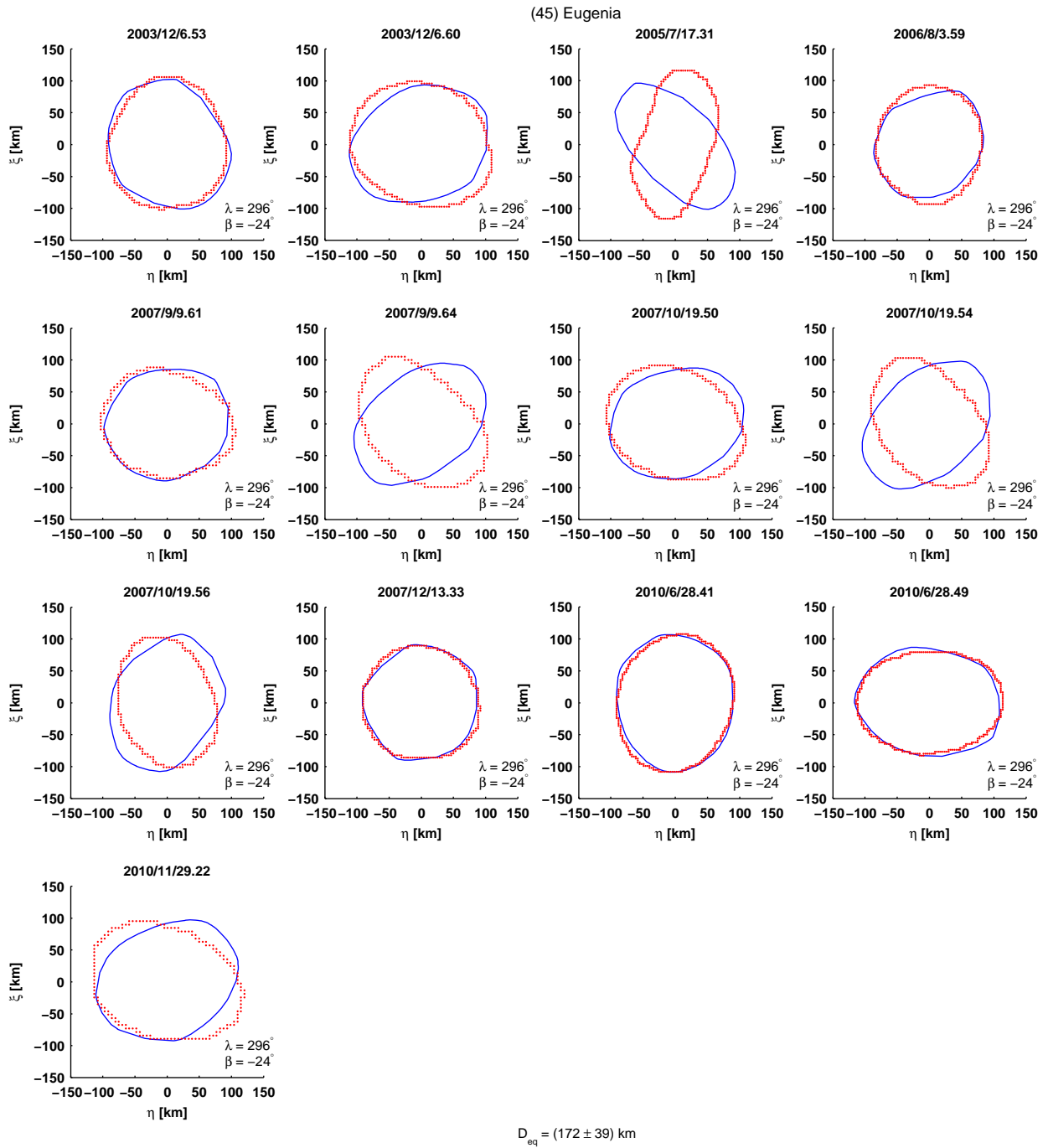}}\\
	 \end{center}
	 \caption{\label{img:45b}(45) Eugenia: Comparison between the AO contours (red dots) and the corresponding convex shape model projections (blue line) of the second pole solution.}
\end{figure}

\newpage

\begin{figure}[!h]
	\begin{center}
	 \resizebox{0.25\hsize}{!}{\includegraphics{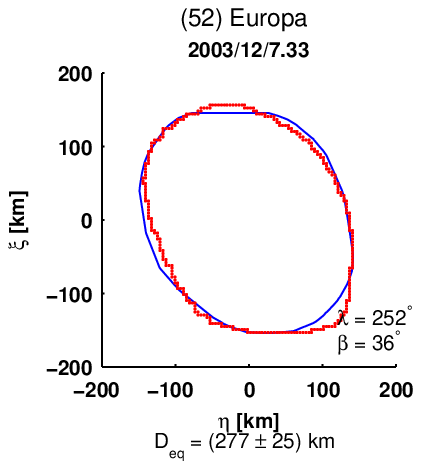}}\\
	 \end{center}
	 \caption{\label{img:52}(52) Europa: Comparison between the AO contour (red dots) and the corresponding convex shape model projection (blue line).}
\end{figure}

\newpage

\begin{figure}[!h]
	\begin{center}
	 \resizebox{1.0\hsize}{!}{\includegraphics{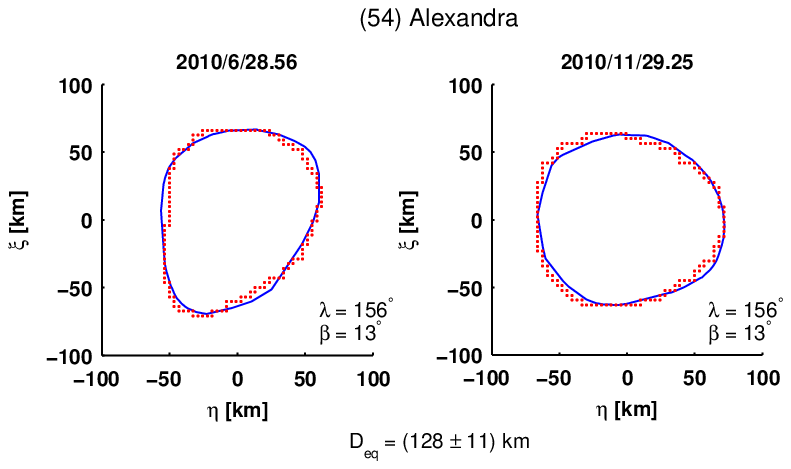}\includegraphics{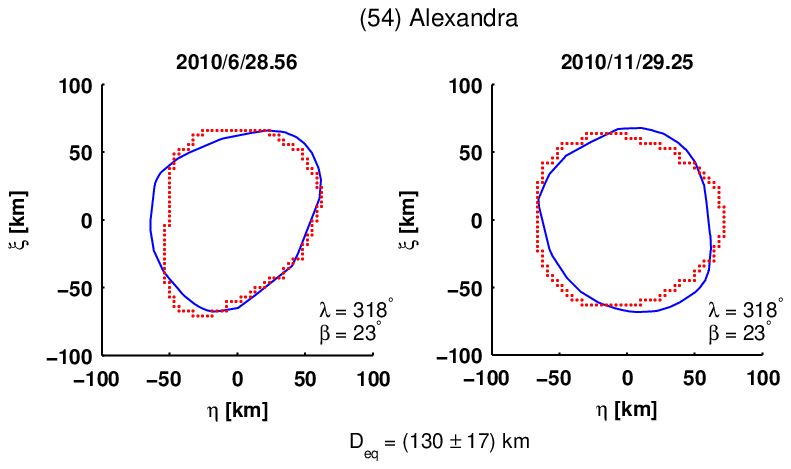}}\\
	 \end{center}
	 \caption{\label{img:54}(54) Alexandra: Comparison between the AO contours (red dots) and the corresponding convex shape model projections (blue line) for both pole solutions. The first pole solution is preferred.}
\end{figure}

\newpage

\begin{figure}[!h]
	\begin{center}
	 \resizebox{0.50\hsize}{!}{\includegraphics{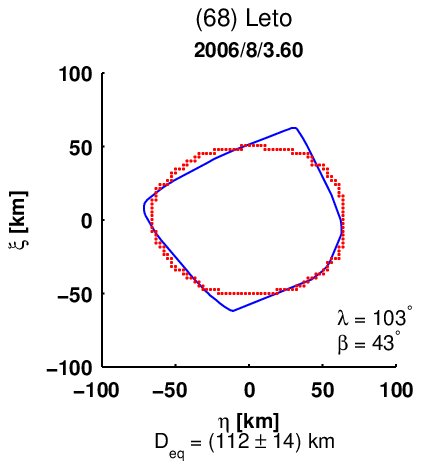}\includegraphics{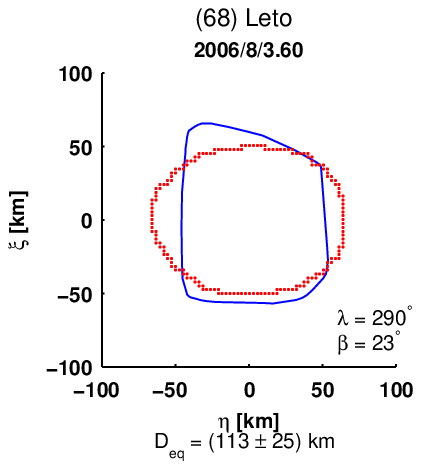}}\\
	 \end{center}
	 \caption{\label{img:68}(68) Leto: Comparison between the AO contour (red dots) and the corresponding convex shape model projection (blue line) for both pole solutions. The first pole solution is preferred.}
\end{figure}

\newpage

\begin{figure}[!h]
	\begin{center}
	 \resizebox{0.50\hsize}{!}{\includegraphics{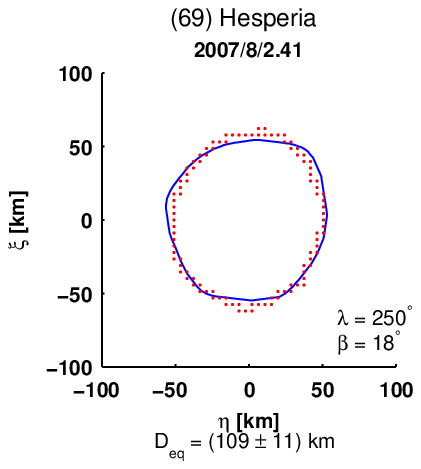}\includegraphics{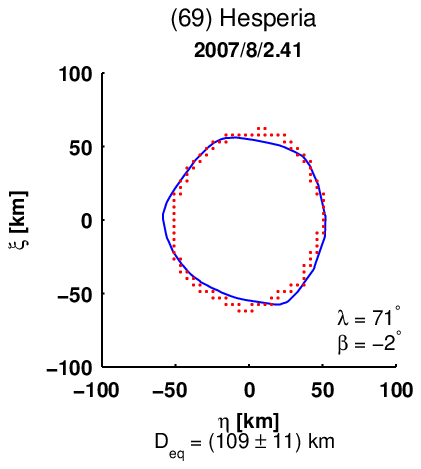}}\\
	 \end{center}
	 \caption{\label{img:69}(69) Hesperia: Comparison between the AO contour (red dots) and the corresponding convex shape model projection (blue line).}
\end{figure}

\newpage

\begin{figure}[!h]
	\begin{center}
	 \resizebox{0.50\hsize}{!}{\includegraphics{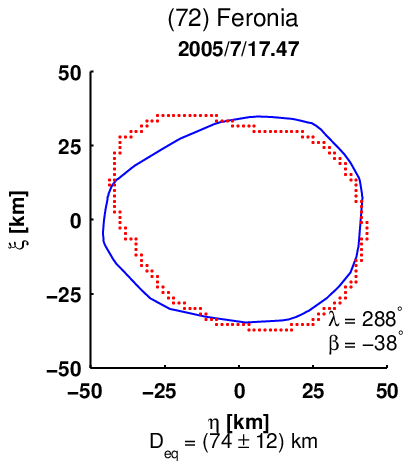}\includegraphics{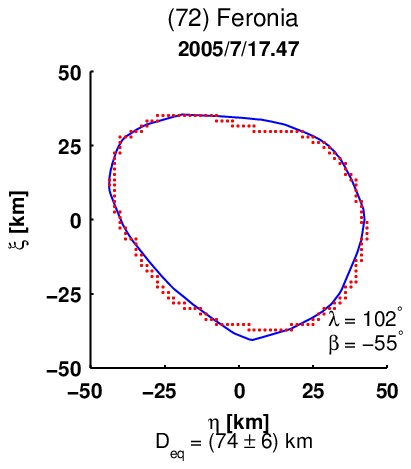}}\\
	 \end{center}
	 \caption{\label{img:72}(72) Feronia: Comparison between the AO contour (red dots) and the corresponding convex shape model projection (blue line) for both pole solutions. The second pole solution is preferred.}
\end{figure}

\newpage
  
\begin{figure}[!h]
	\begin{center}
	 \resizebox{0.50\hsize}{!}{\includegraphics{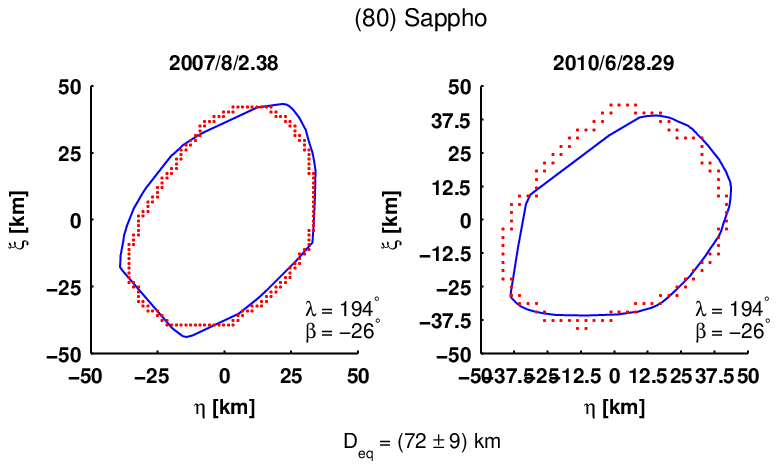}}\\
	 \end{center}
	 \caption{\label{img:80}(80) Sappho: Comparison between the AO contours (red dots) and the corresponding convex shape model projections (blue line).}
\end{figure}

\newpage

\begin{figure}[!h]
	\begin{center}
	 \resizebox{0.25\hsize}{!}{\includegraphics{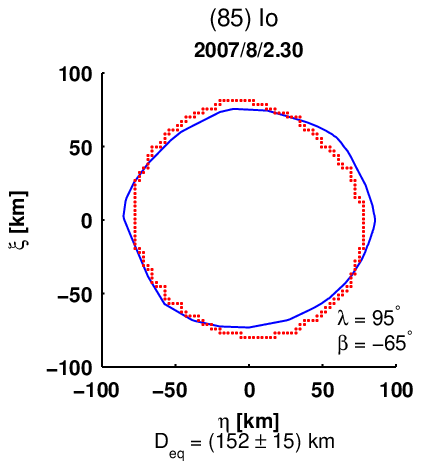}}\\
	 \end{center}
	 \caption{\label{img:85}(85) Io: Comparison between the AO contour (red dots) and the corresponding convex shape model projection (blue line).}
\end{figure}

\newpage

\begin{figure}[!h]
	\begin{center}
	 \resizebox{0.5\hsize}{!}{\includegraphics{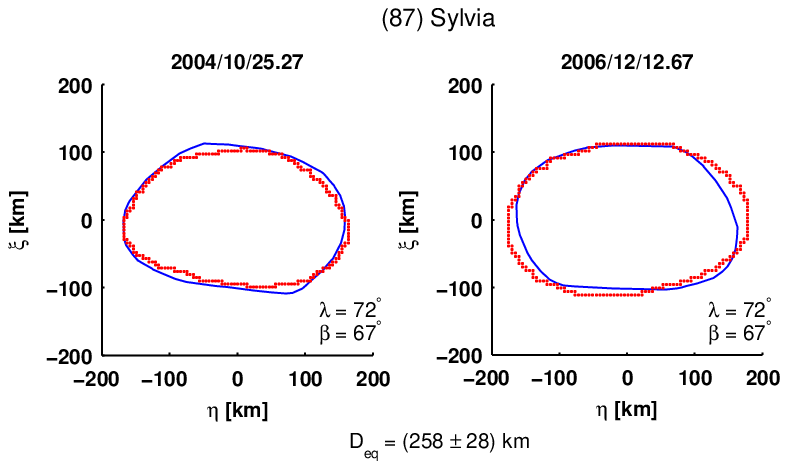}}\\
	 \end{center}
	 \caption{\label{img:87}(87) Sylvia: Comparison between the AO contours (red dots) and the corresponding convex shape model projections (blue line).}
\end{figure}

\newpage

\begin{figure}[!h]
	\begin{center}
	 \resizebox{0.50\hsize}{!}{\includegraphics{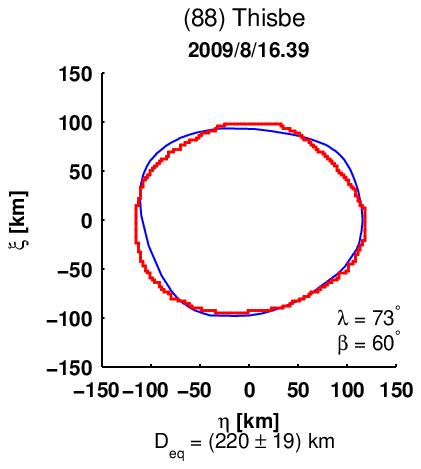}\includegraphics{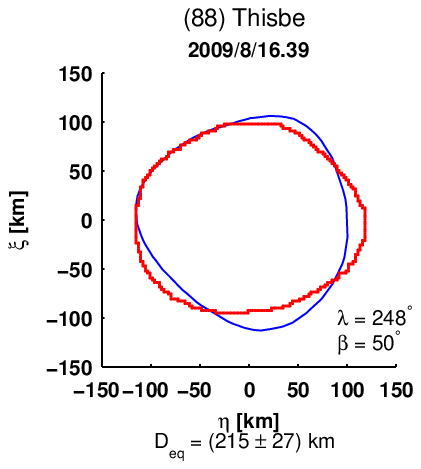}}\\
	 \end{center}
	 \caption{\label{img:88}(88) Thisbe: Comparison between the AO contour (red dots) and the corresponding convex shape model projection (blue line) for both pole solutions. The first pole solution is preferred.}
\end{figure}

\newpage

\begin{figure}[!h]
	\begin{center}
	 \resizebox{0.25\hsize}{!}{\includegraphics{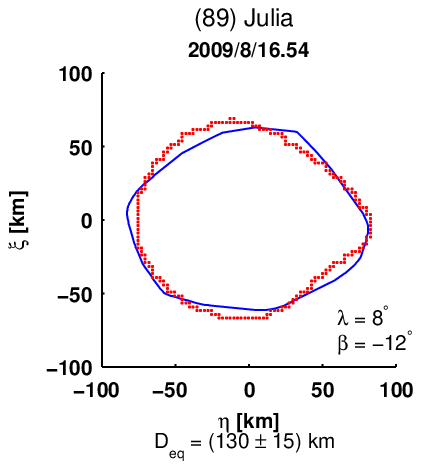}}\\
	 \end{center}
	 \caption{\label{img:89}(89) Julia: Comparison between the AO contour (red dots) and the corresponding convex shape model projection (blue line).}
\end{figure}

\newpage

\begin{figure}[!h]
	\begin{center}
	 \resizebox{0.50\hsize}{!}{\includegraphics{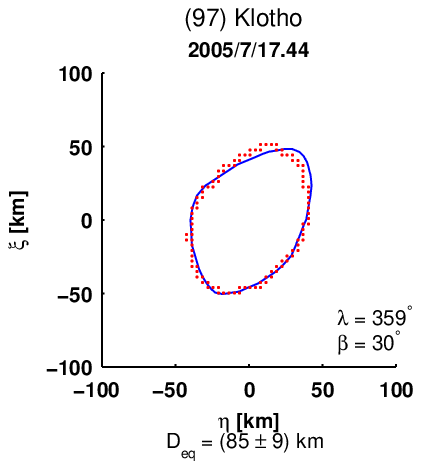}\includegraphics{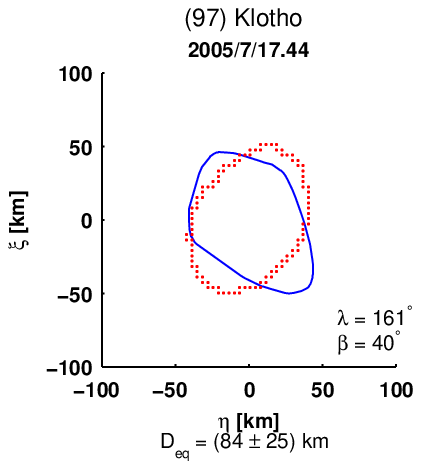}}\\
	 \end{center}
	 \caption{\label{img:97}(97) Thisbe: Comparison between the AO contour (red dots) and the corresponding convex shape model projection (blue line) for both pole solutions. The firt pole solution is preferred.}
\end{figure}

\newpage

\begin{figure}[!h]
	\begin{center}
	 \resizebox{0.75\hsize}{!}{\includegraphics{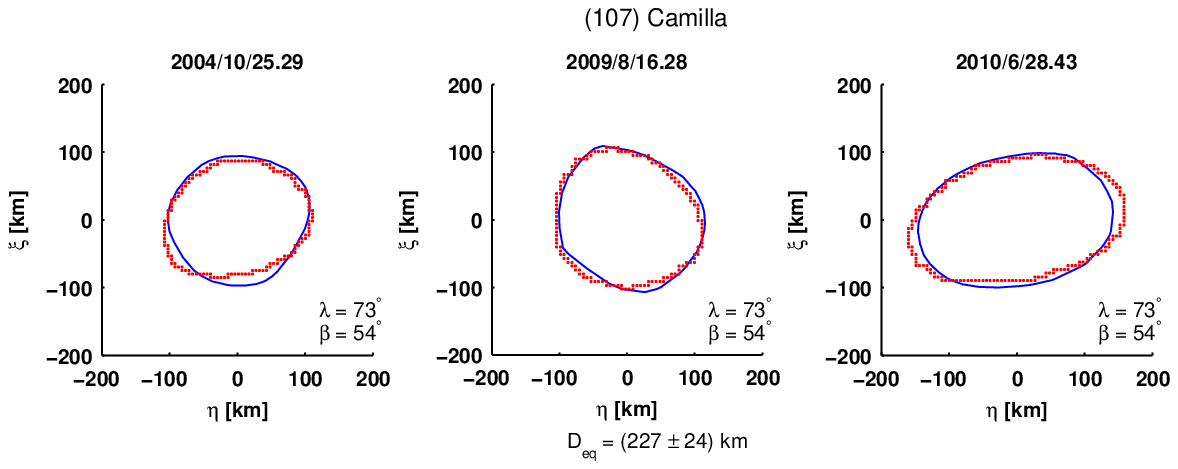}}\\
	 \end{center}
	 \caption{\label{img:107}(107) Camilla: Comparison between the AO contours (red dots) and the corresponding convex shape model projections (blue line).}
\end{figure}

\newpage

\begin{figure}[!h]
	\begin{center}
	 \resizebox{0.25\hsize}{!}{\includegraphics{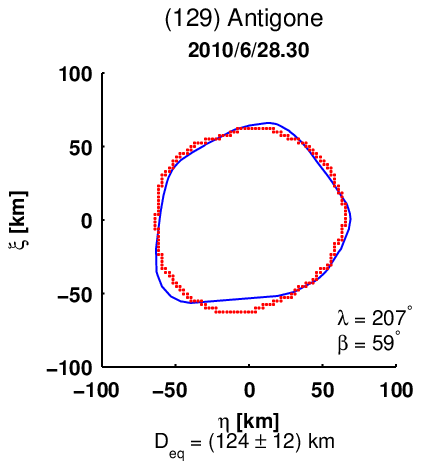}}\\
	 \end{center}
	 \caption{\label{img:129}(129) Antigone: Comparison between the AO contour (red dots) and the corresponding convex shape model projection (blue line).}
\end{figure}

\newpage

\begin{figure}[!h]
	\begin{center}
	 \resizebox{0.75\hsize}{!}{\includegraphics{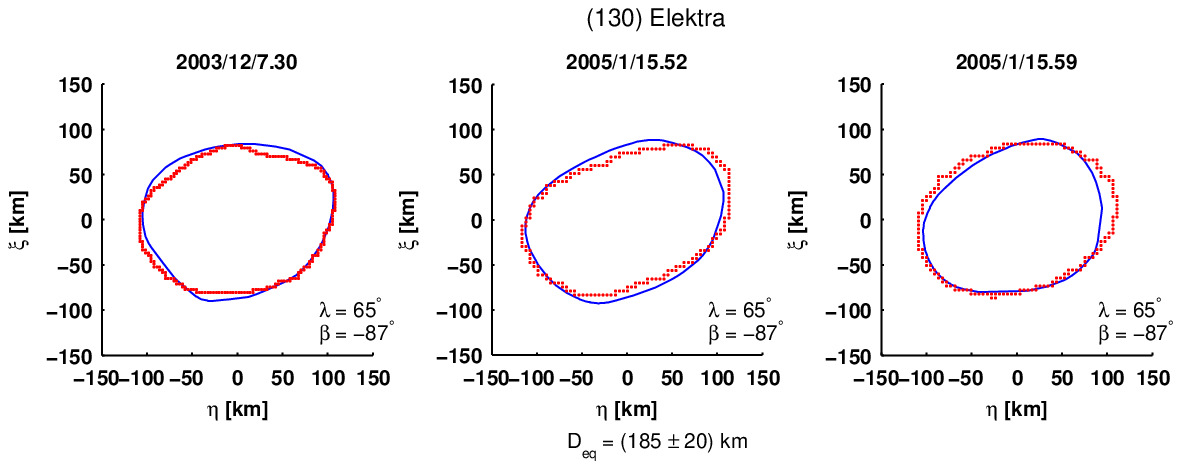}}\\
	 \end{center}
	 \caption{\label{img:130}(130) Elektra: Comparison between the AO contours (red dots) and the corresponding convex shape model projections (blue line).}
\end{figure}

\newpage

\begin{figure}[!h]
	\begin{center}
	 \resizebox{1.0\hsize}{!}{\includegraphics{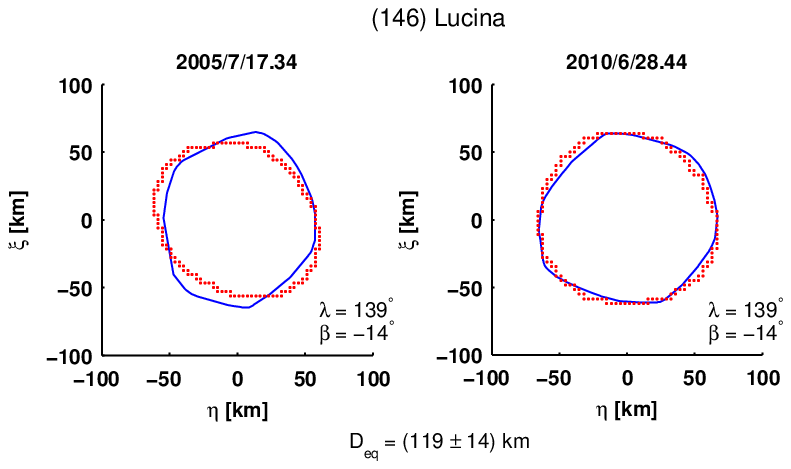}\includegraphics{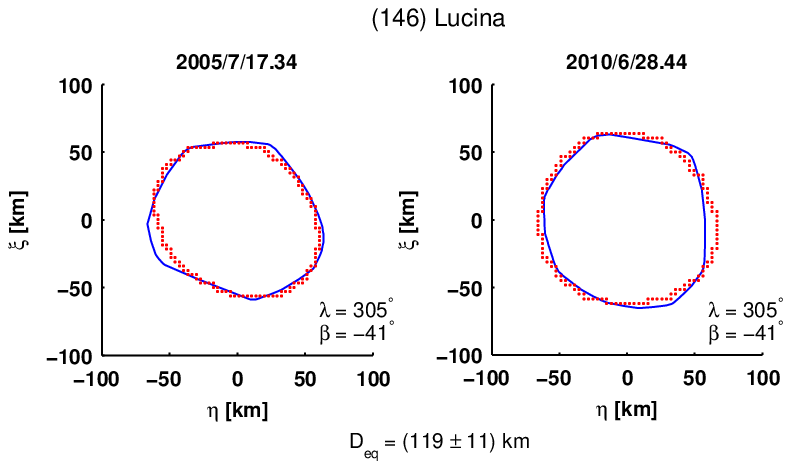}}\\
	 \end{center}
	 \caption{\label{img:146}(146) Lucina: Comparison between the AO contours (red dots) and the corresponding convex shape model projections (blue line) for both pole solutions. The second pole solution is preferred.}
\end{figure}

\newpage

\begin{figure}[!h]
	\begin{center}
	 \resizebox{0.25\hsize}{!}{\includegraphics{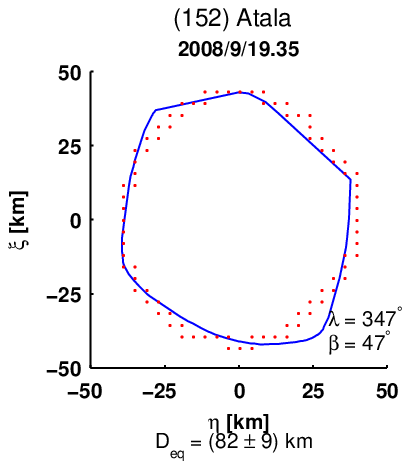}}\\
	 \end{center}
	 \caption{\label{img:152}(152) Atala: Comparison between the AO contour (red dots) and the corresponding convex shape model projection (blue line).}
\end{figure}

\newpage

\begin{figure}[!h]
	\begin{center}
	 \resizebox{0.50\hsize}{!}{\includegraphics{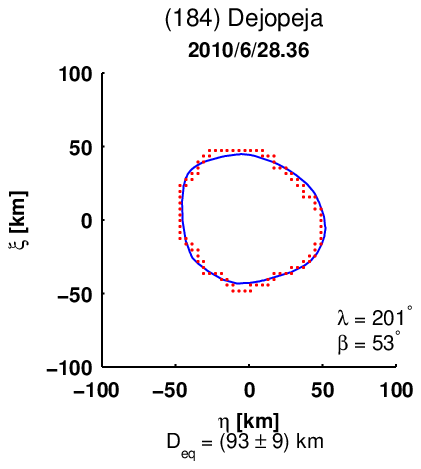}\includegraphics{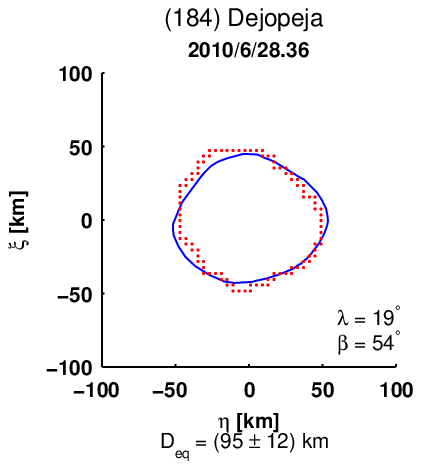}}\\
	 \end{center}
	 \caption{\label{img:184}(184) Dejopeja: Comparison between the AO contour (red dots) and the corresponding convex shape model projection (blue line) for both pole solutions.}
\end{figure}

\newpage

\begin{figure}[!h]
	\begin{center}
	 \resizebox{0.50\hsize}{!}{\includegraphics{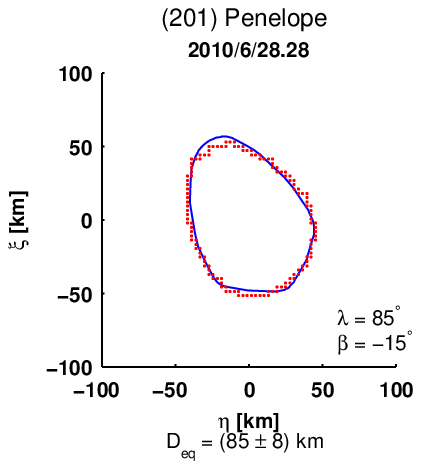}\includegraphics{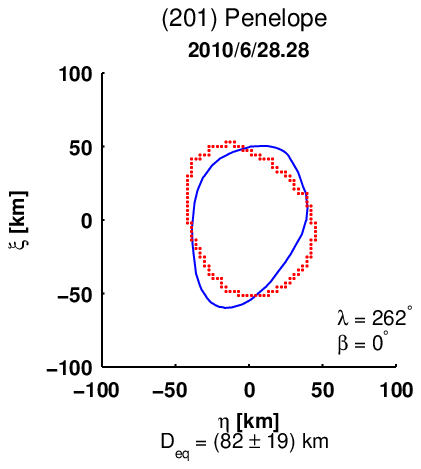}}\\
	 \end{center}
	 \caption{\label{img:201}(201) Penelope: Comparison between the AO contour (red dots) and the corresponding convex shape model projection (blue line) for both pole solutions. The first pole solution is preferred.}
\end{figure}

\newpage

\begin{figure}[!h]
	\begin{center}
	 \resizebox{0.50\hsize}{!}{\includegraphics{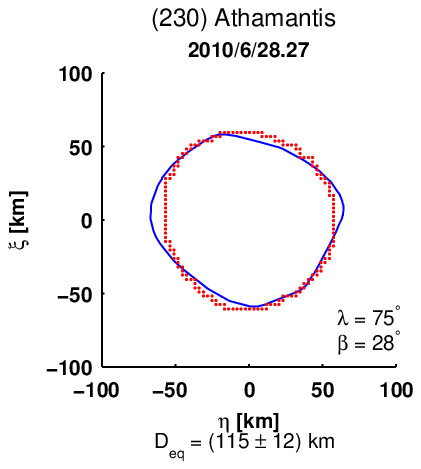}\includegraphics{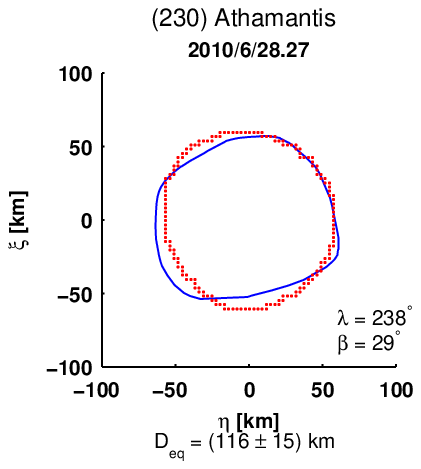}}\\
	 \end{center}
	 \caption{\label{img:230}(230) Athamantis: Comparison between the AO contour (red dots) and the corresponding convex shape model projection (blue line) for both pole solutions.}
\end{figure}

\newpage

\begin{figure}[!h]
	\begin{center}
	 \resizebox{0.50\hsize}{!}{\includegraphics{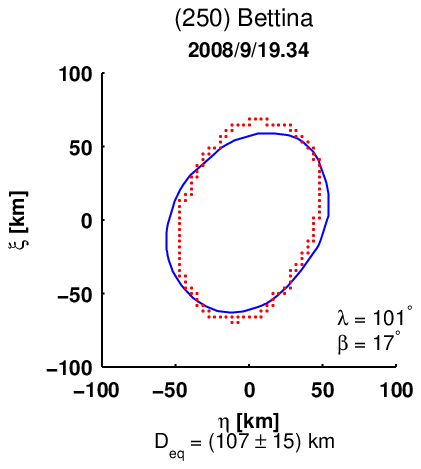}\includegraphics{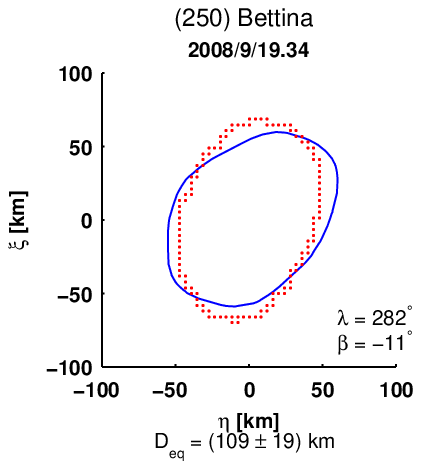}}\\
	 \end{center}
	 \caption{\label{img:250}(250) Bettina: Comparison between the AO contour (red dots) and the corresponding convex shape model projection (blue line) for both pole solutions. The first pole solution is preferred.}
\end{figure}

\newpage

\begin{figure}[!h]
	\begin{center}
	 \resizebox{0.50\hsize}{!}{\includegraphics{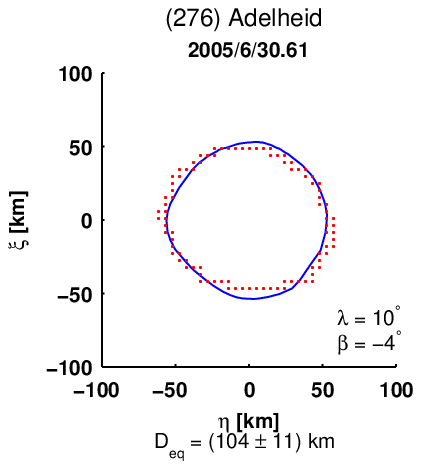}\includegraphics{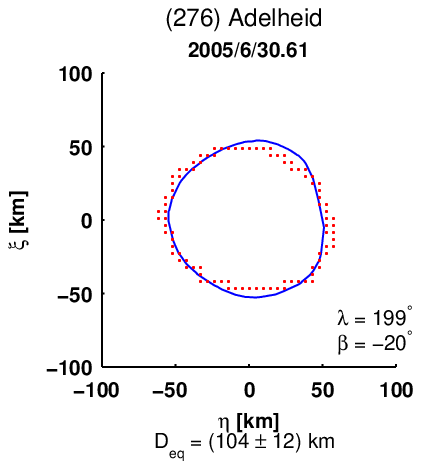}}\\
	 \end{center}
	 \caption{\label{img:276}(276) Adelheid: Comparison between the AO contour (red dots) and the corresponding convex shape model projection (blue line) for both pole solutions.}
\end{figure}

\newpage

\begin{figure}[!h]
	\begin{center}
	 \resizebox{0.50\hsize}{!}{\includegraphics{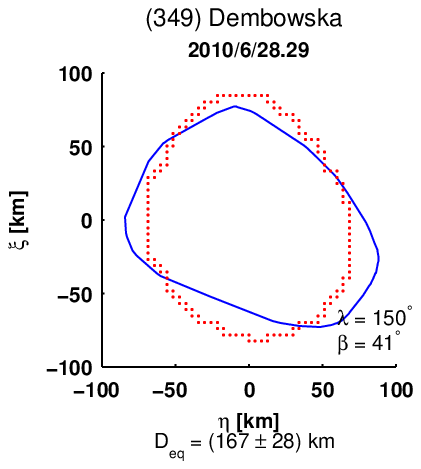}\includegraphics{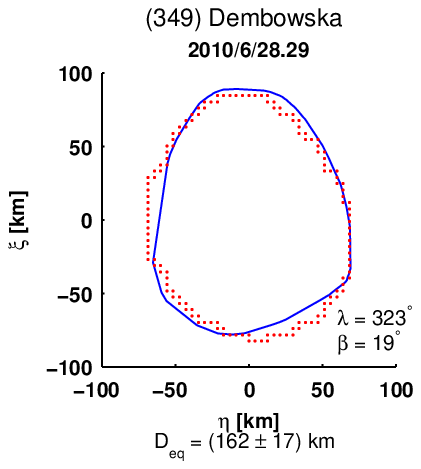}}\\
	 \end{center}
	 \caption{\label{img:349}(349) Dembowska: Comparison between the AO contour (red dots) and the corresponding convex shape model projection (blue line) for both pole solutions. The second pole solution is preferred.}
\end{figure}

\newpage

\begin{figure}[!h]
	\begin{center}
	 \resizebox{0.25\hsize}{!}{\includegraphics{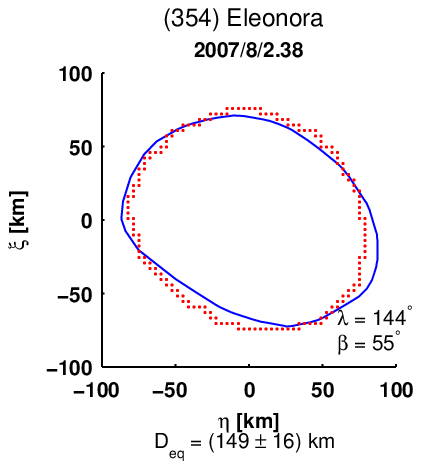}}\\
	 \end{center}
	 \caption{\label{img:354}(354) Antigone: Comparison between the AO contour (red dots) and the corresponding convex shape model projection (blue line).}
\end{figure}

\newpage

\begin{figure}[!h]
	\begin{center}
	 \resizebox{0.25\hsize}{!}{\includegraphics{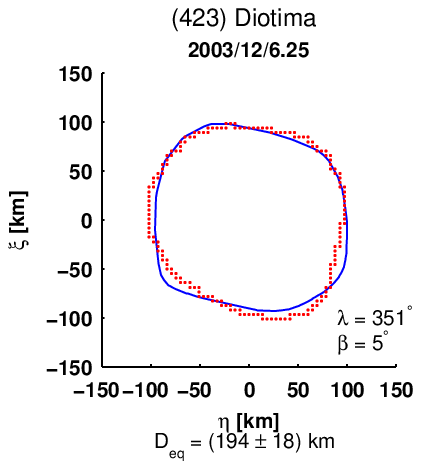}}\\
	 \end{center}
	 \caption{\label{img:423}(423) Diotima: Comparison between the AO contour (red dots) and the corresponding convex shape model projection (blue line).}
\end{figure}

\end{document}